\documentclass[runningheads, hyphens]{llncs}

\usepackage{eccv}

\usepackage[latin9]{inputenc}
\usepackage{array}
\usepackage{multirow}
\usepackage{amsmath}
\usepackage{graphicx}
\usepackage[unicode=true,
 bookmarks=false,
 breaklinks=false,pdfborder={0 0 1},backref=section,colorlinks=false]
 {hyperref}
\hypersetup{
 pagebackref,breaklinks,colorlinks,citecolor=eccvblue}

\makeatletter

\providecommand{\tabularnewline}{\\}

\usepackage{eccvabbrv}

\usepackage[accsupp]{axessibility}

\usepackage{hyperref}

\usepackage{orcidlink}

\usepackage{import}

\usepackage[ruled,vlined]{algorithm2e}
\definecolor{commentcolor}{RGB}{110,154,155}   
\newcommand{\PyComment}[1]{\ttfamily\textcolor{commentcolor}{\# #1}}  
\newcommand{\PyCode}[1]{\ttfamily\textcolor{black}{#1}} 

\makeatother

\begin{document}

\title{PQDynamicISP: Dynamically Controlled Image Signal Processor for Any
Image Sensors Pursuing Perceptual Quality}


\titlerunning{PQDynamicISP}


\author{Masakazu Yoshimura\inst{1} \and
Junji Otsuka\inst{1} \and Takeshi
Ohashi\inst{1}}


\authorrunning{M.~Yoshimura et al.}



\institute{Sony Group Corporation, Tokyo, Japan\\
\email{\{Masakazu.Yoshimura, Junji.Otsuka, Takeshi.A.Ohashi\}@sony.com}}

\maketitle


\begin{abstract}
Full DNN-based image signal processors (ISPs) have been actively studied
and have achieved superior image quality compared to conventional
ISPs. In contrast to this trend, we propose a lightweight ISP that
consists of simple conventional ISP functions but achieves high image
quality by increasing expressiveness. Specifically, instead of tuning
the parameters of the ISP, we propose to control them dynamically
for each environment and even locally. As a result, state-of-the-art
accuracy is achieved on various datasets, including other tasks like
tone mapping and image enhancement, even though ours is lighter than
DNN-based ISPs. Additionally, our method can process different image
sensors with a single ISP through dynamic control, whereas conventional
methods require training for each sensor.  \keywords{Image signal
processor \and image enhancement} 
\end{abstract}

\section{Introduction}

\label{sec:intro}

\begin{figure}[t]
\centering

\begin{tabular}{cc}
\def\svgwidth{0.5\columnwidth}\scriptsize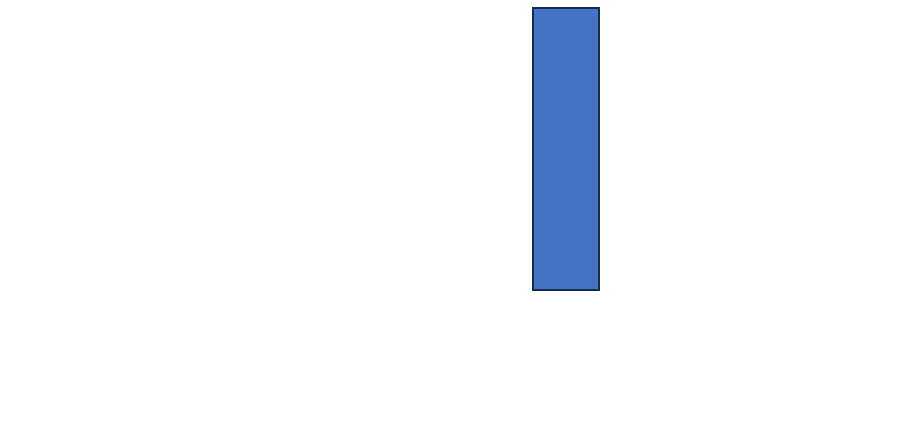 & \def\svgwidth{0.45\columnwidth}\scriptsize
\begingroup%
  \makeatletter%
  \providecommand\color[2][]{%
    \errmessage{(Inkscape) Color is used for the text in Inkscape, but the package 'color.sty' is not loaded}%
    \renewcommand\color[2][]{}%
  }%
  \providecommand\transparent[1]{%
    \errmessage{(Inkscape) Transparency is used (non-zero) for the text in Inkscape, but the package 'transparent.sty' is not loaded}%
    \renewcommand\transparent[1]{}%
  }%
  \providecommand\rotatebox[2]{#2}%
  \newcommand*\fsize{\dimexpr\f@size pt\relax}%
  \newcommand*\lineheight[1]{\fontsize{\fsize}{#1\fsize}\selectfont}%
  \ifx\svgwidth\undefined%
    \setlength{\unitlength}{396.38308548bp}%
    \ifx\svgscale\undefined%
      \relax%
    \else%
      \setlength{\unitlength}{\unitlength * \real{\svgscale}}%
    \fi%
  \else%
    \setlength{\unitlength}{\svgwidth}%
  \fi%
  \global\let\svgwidth\undefined%
  \global\let\svgscale\undefined%
  \makeatother%
  \begin{picture}(1,0.51684821)%
    \lineheight{1}%
    \setlength\tabcolsep{0pt}%
    \put(0,0){\includegraphics[width=\unitlength,page=1]{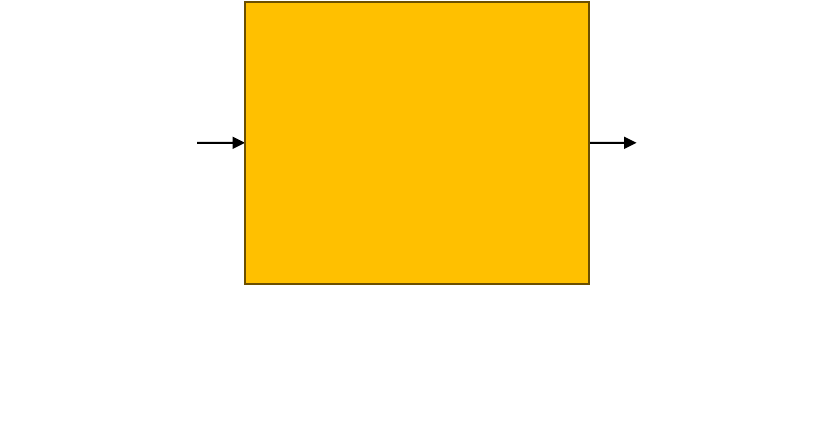}}%
    \put(0.44945899,0.33122325){\color[rgb]{1,1,1}\makebox(0,0)[lt]{\lineheight{1.25}\smash{\begin{tabular}[t]{l}DNN\end{tabular}}}}%
    \put(0,0){\includegraphics[width=\unitlength,page=2]{dnnisp.pdf}}%
    \put(0.02493833,0.1301711){\color[rgb]{0,0,0}\makebox(0,0)[lt]{\lineheight{1.25}\smash{\begin{tabular}[t]{l}i.e. 3024$\times$4032\end{tabular}}}}%
    \put(0,0){\includegraphics[width=\unitlength,page=3]{dnnisp.pdf}}%
  \end{picture}%
\endgroup%
\tabularnewline
\scalebox{0.8}{(a) Conventional ISP} & \scalebox{0.8}{(b) Fully DNN-based ISP}\tabularnewline
\scalebox{0.4}{} & \scalebox{0.4}{}\tabularnewline
\multicolumn{2}{c}{\def\svgwidth{0.65\columnwidth}\scriptsize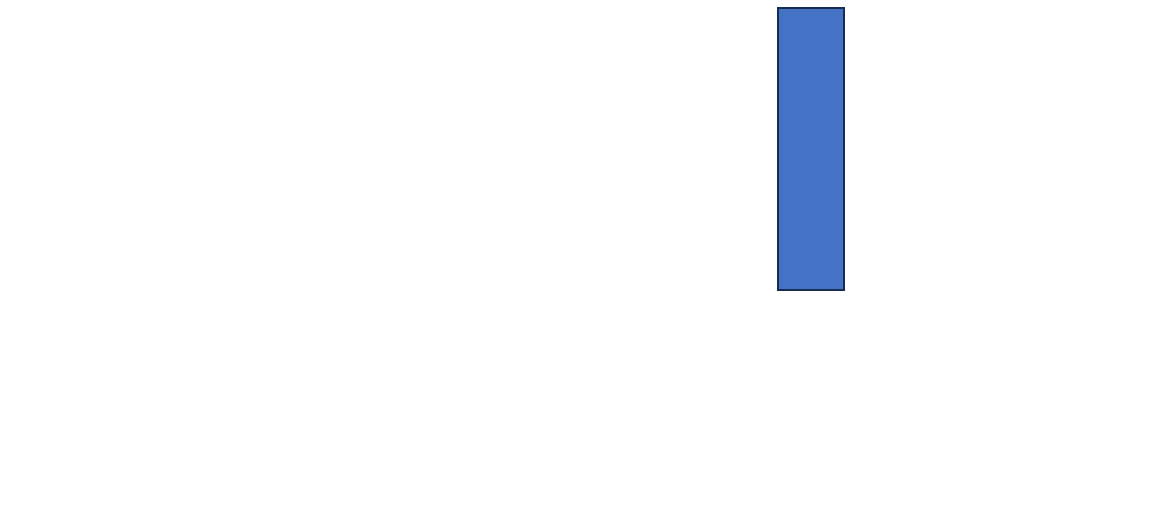}\tabularnewline
\multicolumn{2}{c}{\scalebox{0.8}{(c) Our PQDynacmicISP}}\tabularnewline
\end{tabular}

\caption{(a) The conventional ISPs tune the parameters of classical ISP functions
for each image sensor, while (b) the fully DNN-based ISPs train DNN
for each image sensor. (c) Our method handles any sensors by controlling
the parameters of classical ISP functions for each environment and
each sensor, and even locally.}

\vspace{-4mm}

\label{fig:arch_comparison}
\end{figure}

Image Signal Processors (ISPs) convert RAW images from image sensors
into standard RGB (sRGB) images that appear natural and pleasing to
the human eye. The pixel values in the RAW image are determined by
the physical brightness and image sensor's characteristics. Similarly,
the human eye has its own response characteristics to light, adapting
to dark or light environments and perceiving colors accurately regardless
of ambient light color. Additionally, the relationship between physical
brightness and perceived brightness is non-linear (approximately $y=x^{1/3}$)
\cite{stevens1957psychophysical}. Therefore, ISPs need to cancel
out the sensor characteristics and then mimic the human eye's response
in order to capture the landscape as the human actually perceives
it. Furthermore, some flavoring (\eg, making the sky bluer than it
actually is) may be added \cite{bychkovsky2011learning}.

Conventional ISPs address these challenges by creating and tuning
individual functions, such as demosaicing to increase resolution while
minimizing artifacts like false color and moire, color correction
function to cancel out the sensor's color filter characteristics,
denoising to reduce sensor noise, gain adjustment to mimic the eye's
light/dark adaptation, white balance to mimic color adaptation, and
gamma and tone mapping to mimic the human stimulus response. While
conventional ISPs are lightweight, they lack expressiveness and require
significant time to tune for each image sensor \cite{mosleh2020hardware}.

As a result, there has been extensive research on full Deep Neural
Network (DNN)-based ISPs \cite{dai2020awnet,gou2023syenet,ignatov2022microisp,ignatov2020replacing},
which have demonstrated superior image quality. However, their computational
costs are high because even smartphone cameras tend to require 12MP
($4032\times3024\times3$) resolution images, which is much higher
resolution than the input of typical image recognition DNNs. Also,
existing DNN ISPs are trained for each image sensor because their
characteristics are very different.

While DNN ISPs have been actively studied for perceptual quality,
in the field of image recognition, it has been proven that processing
RAW images with a part of conventional ISP functions before feeding
them into image recognition DNNs can improve recognition accuracy
\cite{yoshimura2023rawgment,mosleh2020hardware,punnappurath2022day,hansen2021isp4ml}.
We infer from this that the data distribution of RAW imagesis challenging
to handle solely with DNNs. Additionally, several methods improve
recognition accuracy even more by dynamically controlling the parameters
of the classical ISP functions to make it easier for the recognition
DNN to recognize \cite{yoshimura2023dynamicisp,qin2023learning,qin2022attention,liu2023improving}.

Therefore, we challenge the realization of a lightweight and high-quality
ISP by dynamically controlling the parameters of conventional ISPs
for perceptual image quality. For image recognition purposes, it is
sufficient to control a portion of the ISP functions since downstream
recognition DNNs have the ability to process any data distributions
to some extent. Even if the control fails, it is acceptable as long
as the recognition DNN recognizes the image correctly \cite{liu2023improving,yoshimura2023dynamicisp}.
However, when it comes to the ISP for perceptual image quality, it
is crucial to successfully address the cancellation of sensor characteristics,
adaptation to human eye characteristics, and flavoring without failure.
Another problem is that classical ISP functions have many local minima
\cite{tseng2019hyperparameter}. In this work, we propose training
methods to tackle these challenges. Furthermore, to enhance performance,
we make the ISP parameters controllable not only frame by frame but
also locally. As a result, our work outperforms even large-scale NNs.

The mechanism of the dynamic ISP control has the potential to create
a universal ISP that can handle various image sensors. It can predict
sensor characteristics based on input images and automatically determine
the appropriate parameters for each sensor and environment. We introduce
a new task in which RAW images from various image sensors are processed
without any distinctions. Our results demonstrate significant advantages
over existing works. A single ISP that can be used across different
image sensors would be highly beneficial, eliminating the need for
data collection and tuning for each individual sensor.

In summary, the contributions of this work are:
\begin{itemize}
\item Dynamic control of classical ISP functions for image quality is proposed
and achieves state-of-the-art image quality that surpasses large DNN
ISPs.
\item Our proposed method not only controls ISP parameters frame by frame
but also locally, resulting in further improvements in accuracy.
\item Training methods to deal with the local minima in the ISP are proposed. 
\item We newly define an universal ISP task and demonstrate that our method
offers significant advantages over existing methods.
\item Furthermore, we repurpose our ISP for tone mapping and enhancement
tasks, showcasing its versatility.
\end{itemize}

\section{Related Works}

\subsection{ISP}

As mentioned above, ISPs have various roles, so in the early era,
each function such as white balance \cite{van2007edge,cheng2014illuminant},
denoiser \cite{dabov2007image,buades2005non}, and tone mapping \cite{shibata2016gradient,gu2012local,drago2003adaptive},
evolved as separate algorithms. Even now, improving each algorithm
of white balance \cite{afifi2021cross,ershov2023physically}, denoiser
\cite{li2023spatially,zou2023iterative}, and tone mapping \cite{li2023spatially,yang2022adaint,yang2022seplut}
using DNNs is an important research topic. On the other hand, research
on end-to-end ISP parameter tuning has also emerged to improve perceptual
image quality \cite{mosleh2020hardware,pavithra2021automatic,hevia2020optimization,wu2019visionisp}
and downstream recognition accuracy \cite{mosleh2020hardware,yoshimura2023rawgment,onzon2021neural}.

In recent years, full DNN-based ISPs have been actively studied. PyNet
\cite{ignatov2020replacing} and AWNet \cite{dai2020awnet} propose
high-quality DNN-based ISPs. More recently, many efforts have been
made to develop practical lightweight DNN ISPs for smartphones and
DSLR cameras \cite{gou2023syenet,ignatov2022microisp,zhang2021learning}. 

However, ISPs have various roles, so the difficulty of replacing an
ISP with a single DNN is becoming apparent. For example, CameraNet
\cite{liang2021cameranet} shows significant improvement in image
quality by splitting the ISP into two DNNs: an image restoration DNN
and a color manipulation DNN. Deep-FlexISP \cite{liu2022deep} won
the NTIRE 2022 challenge \cite{ershov2022ntire} by further splitting
it into three DNNs. In the following year, a method that carefully
tunes classical ISP functions \cite{zini2023back} won the NTIRE 2023
challenge \cite{shutova2023ntire}, suggesting that classical ISP
functions are still competitive depending on the quality and quantity
of the dataset.

\subsection{Dynamic Control and Local Parameter Tuning of ISP}

Dynamic ISP control to improve downstream image recognition has started
to emerge. Two works control a classical elaborate black box ISP \cite{qin2022attention,qin2023learning}
for a frozen downstream detector. They also show potential for improving
image quality in a toy task. K. Q. Dinh \etal \cite{dinh2023end}
and DynamicISP \cite{yoshimura2023dynamicisp} achieve significant
accuracy improvement by controlling several simple differentiable
ISP functions and optimizing the ISP and downstream model end-to-end.
Several inverse ISP methods also control the parameters of inverse
ISP functions \cite{conde2022model,otsuka2023self}.

For the tone mapping task, which is part of the ISP task, many methods
achieve high image quality by generating a 3D look-up table for each
image \cite{yang2022seplut,yang2022adaint,zeng2020learning}. These
are also a type of dynamic ISP control, but the cost of practical
use in small devices is high due to high memory consumption \cite{conde2023nilut}
and lack of backend implementations, such as ONNX Runtime \cite{onnxruntime}
and TensorFlow Lite \cite{tflite}. Therefore, in this study, we only
control the parameters of simple functions.

In addition, local tone mapping, which tunes ISP parameters locally
based on local luminance distribution, has improved visibility in
HDR environments \cite{shibata2016gradient,gu2012local}. Based on
the local tone mapping idea, we propose a method to control ISP parameters
not only frame by frame but also locally.

\section{Methodology}

As mentioned above, we propose dynamic ISP parameter control for perceptual
image quality. The basic idea is based on DynamicISP \cite{yoshimura2023dynamicisp},
a method used for enhancing image recognition. We need to propose
a new method because it requires finer control to achieve high image
quality that feels natural to people, rather than just being able
to distinguish what an object is. The proposed model structure and
its training method are described below. The model structure can be
divided into three main parts: the ISP, the lightweight encoder, and
the controller that determines the ISP parameters, as shown in Fig.
\ref{fig:arch_comparison}(c).

\subsection{ISP}

Conventional ISPs used in smartphones and SLR cameras are not as redundant
as DNNs, but they are complex with many hyper-parameters. It was found
that the computing speed is slow despite their FLOPs due to the lack
of optimization in deep learning frameworks. Therefore, we propose
an ISP that is as lightweight as possible, while having generality
that can be used for various datasets. The proposed ISP consists of
only five elements: denoiser $I_{DN}$, color correction $I_{CC}$,
gain $I_{GA}$, tone mapping $I_{TM}$, and contrast stretcher $I_{CS}$.
The input image $X$ is sequentially processed by them like, $I_{CS}(I_{TM}(I_{GA}(I_{CC}(I_{DS}(X)))))$.
We define the input and output range of the ISP as {[}0, 1{]} in the
following formulas.

\subsubsection*{Denoiser.}

While DynamicISP uses a simple Gaussian denoiser, DNN-based denoisers
are powerful. In pursuit of perceptual quality, other filtering processes
such as demosaicing, blur removal, and sharpening are also necessary,
and a single DNN can encompass multiple filtering processes. Therefore,
a DNN is used as a denoiser in this work. However, the computational
cost of even a few-layer DNNs is enormous when processing high-resolution
images.

\begin{figure}
\vspace{-2mm}

\centering

\def\svgwidth{0.63\columnwidth}

\scriptsize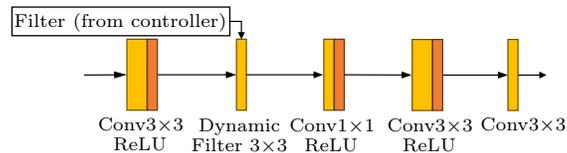

\caption{The proposed light weight denoiser.}

\vspace{-8mm}

\label{fig:denoiser}
\end{figure}

Therefore, we propose a denoiser built with a tiny CNN, but some of
the convolution kernels are dynamically generated to increase the
expressive power as shown in Fig. \ref{fig:denoiser}. Specifically,
a depth-wise convolution kernel is generated for each image: $I_{Dy}(X)=Conv(P_{filter},X)$,
where $P_{filter}$ is a depth-wise convolution kernel of $C_{in}\times1\times k\times k$
size. If the local ISP control proposed later is used, it becomes
a depth-wise dynamic filter of $H\times W\times C_{in}\times1\times k\times k$
size. Conventional dynamic filter methods generate filters from the
previous intermediate features \cite{zhou2021decoupled,shen2023adaptive},
so they are necessary to encode high-resolution features. However,
our method uses the controller described later to generate filters,
which successfully reduces the computational cost.

\subsubsection*{Color Correction Including White Balance.}

The general color correction matrix cancels the color filter characteristic
of the sensor and is a fixed $3\times3$ matrix. Our method, on the
other hand, controls nine parameters of a $3\times3$ matrix $P_{CC}$
in the following function; $I_{CC}\left(X\right)=XP_{CC}$. In usual
ISPs, white balance is also used as a separate module from color correction
to imitate the human eye's color adaptation characteristics. However,
to reduce the computational cost, we make our color correction function
simultaneously cancel the color filter characteristics and imitate
the color adaptation.

\subsubsection*{Gain.}

Our gain function is based on DynmaicISP's implementation, which amplifies
image values while avoiding overflow from $[0,1]$ without clipping.
We improve the expressive power without increasing the computational
cost by using different parameters for each color channel, $c=\{r,g,b\}$,
as follows; 

\begin{equation}
I_{GA}\left(X_{c}\right)=\begin{cases}
\frac{1-p_{h,c}}{1-p_{w,c}}X_{c}\;\left(if\,x_{c}<p_{x,c}\left(1-p_{w,c}\right),\,x_{c}\in X_{c}\right) & \,\\
\frac{1-p_{h,c}}{1-p_{w,c}}X_{c}+\frac{p_{h,c}-p_{w,c}}{1-p_{w,c}}\;\left(if\,p_{x,c}\left(1-p_{w,c}\right)+p_{w,c}<x_{c}\right) & \,\\
\frac{p_{h,c}}{p_{w,c}}(X_{c}-p_{x,c}(1-p_{w,c}))+p_{x,c}(1-p_{h,c})\;\left(otherwise\right) & \,
\end{cases},\label{eq:ga}
\end{equation}
 where $X_{c}$ is each channel array of $X$. We further improve
the function by decorating $p_{x,c}$ as $p_{x,c}=10^{p'_{x,c}}$
to control dark areas more precisely, as the dark value is stretched
in the subsequent module of tone mapping.

\subsubsection*{Tone Mapping.}

We follow the parameterization of the previous gamma tone mapping
\cite{mosleh2020hardware,yoshimura2023rawgment,yoshimura2023dynamicisp}
and further parameterize it by using different parameters for each
color channel as follows;

\begin{equation}
I_{TM}\left(X_{c}\right)=X_{c}^{\frac{1}{p_{\text{\ensuremath{\gamma}1,c}}}\cdot\frac{1-\left(1-p_{\text{\ensuremath{\gamma}2,c}}\right)X_{c}^{\frac{1}{p_{\text{\ensuremath{\gamma}1,c}}}}}{1-\left(1-p_{\text{\ensuremath{\gamma}2,c}}\right)p_{k,c}^{\frac{1}{\text{\ensuremath{p_{\text{\ensuremath{\gamma}1,c}}}}}}}}.\label{eq:gm}
\end{equation}

\subsubsection*{Contrast Stretcher.}

As the recognizer can process values other than {[}0, 1{]}, DynamicISP
simply uses the linear function $Y=p_{a}X+p_{b}$. However, for perceptual
image quality, it is necessary to keep the value within {[}0, 1{]}.
So eq. \ref{eq:ga} is used again but, as it is after tone mapping,
the further decoration is not used.

\subsubsection*{Inverse Tone Mapping.}

We attempt to apply our work not only to the ISP task but also to
the enhancement task, which converts sRGB images to clean sRGB images.
In this case, an inverse tone mapping function is added at the beginning
of our ISP to convert the original sRGB to linear sRGB. We construct
the following function to be similar to the inverse function of eq.
\ref{eq:gm}: 
\begin{equation}
I_{IT}(X_{c})=X_{c}^{p_{\text{\ensuremath{\gamma}3,c}}\cdot\frac{1+p_{\text{\ensuremath{\gamma}4,c}}(X_{c}+1)^{p_{\text{\ensuremath{\gamma}3,c}}}}{1+p_{\text{\ensuremath{\gamma}4,c}}(p_{k2,c}+1)^{p_{\text{\ensuremath{\gamma}3,c}}}}}.\label{eq:it}
\end{equation}

\subsection{Encoder}

\begin{figure}[t]
\centering

\begin{tabular}{cc}
\def\svgwidth{0.25\columnwidth}\scriptsize
\begingroup%
  \makeatletter%
  \providecommand\color[2][]{%
    \errmessage{(Inkscape) Color is used for the text in Inkscape, but the package 'color.sty' is not loaded}%
    \renewcommand\color[2][]{}%
  }%
  \providecommand\transparent[1]{%
    \errmessage{(Inkscape) Transparency is used (non-zero) for the text in Inkscape, but the package 'transparent.sty' is not loaded}%
    \renewcommand\transparent[1]{}%
  }%
  \providecommand\rotatebox[2]{#2}%
  \newcommand*\fsize{\dimexpr\f@size pt\relax}%
  \newcommand*\lineheight[1]{\fontsize{\fsize}{#1\fsize}\selectfont}%
  \ifx\svgwidth\undefined%
    \setlength{\unitlength}{251.26504493bp}%
    \ifx\svgscale\undefined%
      \relax%
    \else%
      \setlength{\unitlength}{\unitlength * \real{\svgscale}}%
    \fi%
  \else%
    \setlength{\unitlength}{\svgwidth}%
  \fi%
  \global\let\svgwidth\undefined%
  \global\let\svgscale\undefined%
  \makeatother%
  \begin{picture}(1,0.77082447)%
    \lineheight{1}%
    \setlength\tabcolsep{0pt}%
    \put(0,0){\includegraphics[width=\unitlength,page=1]{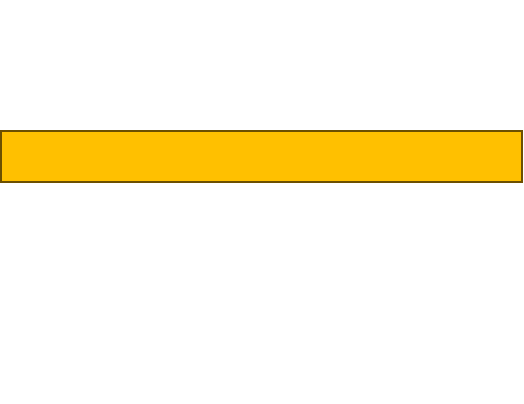}}%
    \put(0.26452156,0.45152855){\color[rgb]{1,1,1}\makebox(0,0)[lt]{\lineheight{1.25}\smash{\begin{tabular}[t]{l}block ($k=5$)\end{tabular}}}}%
    \put(0,0){\includegraphics[width=\unitlength,page=2]{encoder_overview.pdf}}%
    \put(0.17423818,0.63085267){\color[rgb]{1,1,1}\makebox(0,0)[lt]{\lineheight{1.25}\smash{\begin{tabular}[t]{l}Resize (224$\times$224)\end{tabular}}}}%
    \put(0,0){\includegraphics[width=\unitlength,page=3]{encoder_overview.pdf}}%
    \put(0.26429854,0.27454413){\color[rgb]{1,1,1}\makebox(0,0)[lt]{\lineheight{1.25}\smash{\begin{tabular}[t]{l}block($k=3$)\end{tabular}}}}%
    \put(0,0){\includegraphics[width=\unitlength,page=4]{encoder_overview.pdf}}%
    \put(0.25827316,0.09748011){\color[rgb]{1,1,1}\makebox(0,0)[lt]{\lineheight{1.25}\smash{\begin{tabular}[t]{l}block($k=3$)\end{tabular}}}}%
    \put(0,0){\includegraphics[width=\unitlength,page=5]{encoder_overview.pdf}}%
  \end{picture}%
\endgroup%
 & \def\svgwidth{0.72\columnwidth}\scriptsize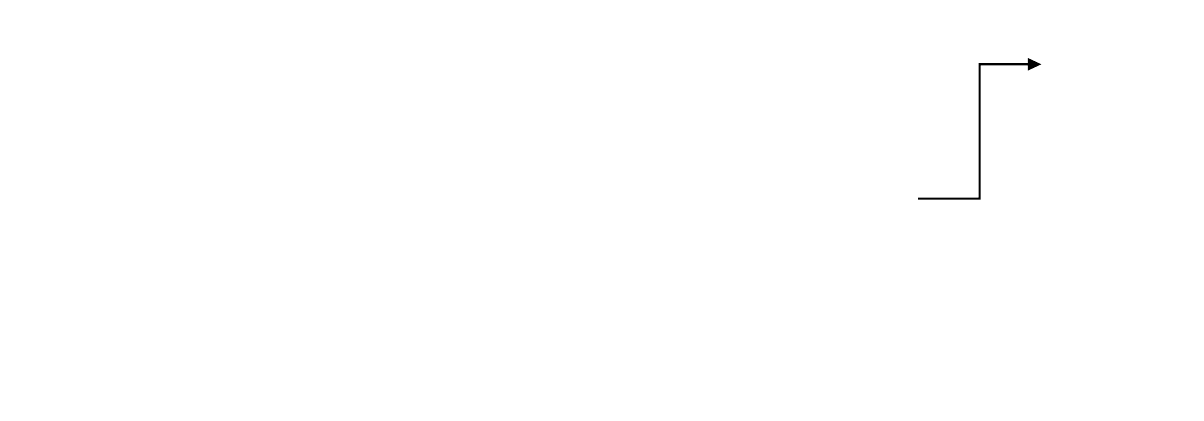\tabularnewline
\scalebox{0.8}{(a) Proposed encoder} & \scalebox{0.8}{(b) The detail of each encoder block }\tabularnewline
\end{tabular}

\caption{The proposed (a) light weight encoder consists of three (b) blocks
modified from the SYENet block \cite{gou2023syenet}.}

\vspace{-2mm}

\label{fig:encoder}
\end{figure}

In DynamicISP, an intermediate feature of a recognizer is used as
an information source to determine the ISP parameters. However, ISP
control for perceptual image quality requires an image encoder to
obtain the feature and determine the ISP parameters. Therefore, a
lightweight encoder is proposed as shown in Fig. \ref{fig:encoder}(a).
First, any resolution of inputs is resized to a fixed resolution in
order to reduce computational costs. Since the SYENet block achieved
high performance in image restoration with only a few layers of CNNs
\cite{gou2023syenet}, we assumed that the SYENet block could generate
a good feature with a few layers. However, unlike when using the SYENet
block for image restoration, it was found that the training loss exploded
when using it as a feature extractor, as there are fewer constraints
on the output of the encoder. Therefore, layer normalization is added
as shown in Fig. \ref{fig:encoder}(b). In addition, stride convolution
is introduced to the first layer of each block to reduce computational
costs significantly.

\subsection{Controller}

Our controller controls the ISP parameters based on the feature from
the encoder. We propose two controllers: a lightweight global controller
and a more precise local controller.

\subsubsection*{Global Control.}

In the global control, the ISP parameters are controlled per image.
Therefore, global average pooling is applied to the encoder's output
to create global latent variable $V_{0}$, and the parameters are
determined with a fully connected layer $f_{full,l}$ as follows:
\begin{equation}
P_{l}=f_{dec,l}(V_{l-1})=f_{act,l}\left(\hat{P_{l}}+f_{full,l}(V_{l-1})\right),\label{eq:dec}
\end{equation}
 where $V_{l-1}$ is the latent variable to decide ISP parameters
$P_{l}$ in the $l$-th ISP layer, and $\hat{P_{l}}$ are learnable
parameters. Following DynamicISP, we define $f_{act,l}$ as 
\begin{equation}
f_{act,l}(x)=\left(P_{l,max}-P_{l,min}\right)\cdot sigmoid(x)+P_{l,min},\label{eq:4}
\end{equation}
 to control parameters within $\left(P_{l,min},P_{l,max}\right)$.
We update the latent variable considering what ISP parameters are
used in each layer by $V_{l}=f_{up,l}\left(V_{l-1},P_{l}\right)$
the samely with DynamicISP but we use proposed group-wise sigmoid
cross-attention as $f_{up,l}$ to achieve precise control. It is different
from usual cross-attention \cite{rombach2022high} in six points as
shown in \cref{alg:sigattn}; (1) linear layers are only used for
generating the key, (2) the content input ($P_{l}$) is not used as
the source of the value, (3) sigmoid is used instead of softmax for
the attention, (4) the last matrix multiplication with the value is
replaced with a Hadamard product, (5) the last feed forward network
is not used, and (6) a virtual sequential length is introduced. Unlike
vision transformers, the sequential length of the input is one. So,
we group the channel dimension of the latent variable $V_{l-1}$ with
the virtual sequential length and attend within the groups. The motivation
for the proposal is as follows. If normal cross-attention is used,
the distribution of the query ($V_{l-1}$) is mixed according to the
key ($P_{l}$) and value ($P_{l}$). However, in the current problem
set-up, the sources of the key ($P_{l}$) and value ($P_{l}$) are
a few parameters, and they don't have enough information to mix the
high-dimensional $V_{l-1}$ appropriately. Therefore, not to corrupt
the distribution of $V_{l-1}$, our method determines how much each
value of $V_{l-1}$ should be changed considering $P_{l}$ and $V_{l-1}$
of the same group, and just multiplies to $V_{l-1}$ using a simple
Hadamard product.

\begin{algorithm}[tb] 
\caption{Group-wise Sigmoid Cross-Attention}\label{alg:sigattn} 
\SetAlgoLined
\PyCode{def \_\_init\_\_(C, Num\_param, vertual\_seq):} \\
	\Indp
	\PyCode{emb\_dim = C/vertual\_seq} \\
	\PyCode{linears = Sequential(Linear(), ReLU(), Linear(c\_out=emb\_dim**2))}\\ 
	\PyCode{scale = emb\_dim ** -0.5} \\
	\Indm
\PyCode{def forward($V_{l-1}$, $P_l$):} \PyComment{[B,C], [B, ($l$-th ISP layer's) Num\_param]}\\
	\Indp
	\PyCode{q = $V_{l-1}$.reshape(B, vertual\_seq, emb\_dim)} \\
	\PyCode{k = linears(($P_l$ - $P_{l,min}$) / ($P_{l,max}$ - $P_{l,min}$))} \\
	\PyCode{k = k.reshape(B, emb\_dim, emb\_dim)} \\
    \PyCode{v = $V_{l-1}$} \\
	\PyCode{attn = 5 * sigmoid(q@k * scale).view(B, -1)} \\
    \PyCode{$V_l$ = v*attn} \\
	\PyCode{return $V_l$} \\
	\Indm
\end{algorithm}

\subsubsection*{Local Control.}

\begin{figure}[t]
\centering

\def\svgwidth{1.0\columnwidth}

\scriptsize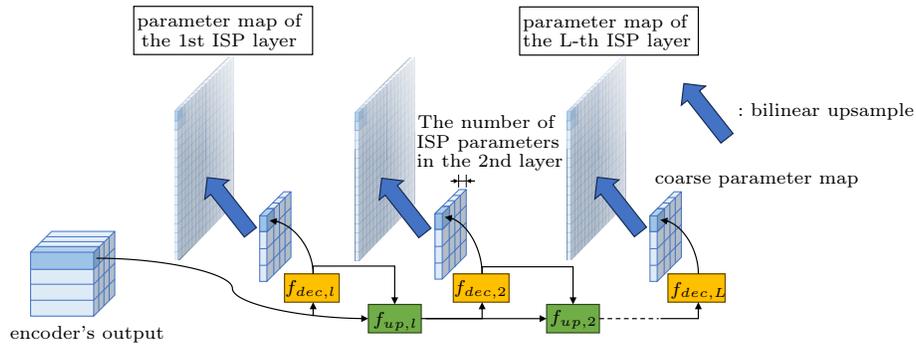

\caption{The proposed local controller. In practical implementation, we don't
iterate per region but compute at once using point-wise convolutions
within $f_{dec,l}$ and $f_{up,l}$.}

\label{fig:controller}
\end{figure}

In local control, the ISP parameters are controlled on a pixel-by-pixel
basis in addition to an image-by-image basis. However, if the control
is completely per-pixel, the computational cost of generating ISP
parameters is high and the accuracy is not good due to too many degrees
of freedom. Therefore, the ISP parameters are generated at the coarse
resolution of the encoder output, which is then expanded to the original
image size by bilinear upsampling. In other words, the global average
pooling for global control is removed, the parameters are generated
for each region of the encoder's output resolution in the same way
as global control, and finally bilinear upsampling is applied as shown
in Fig. \ref{fig:controller}.

\subsection{Training Methods}

\subsubsection*{Automatic Tuning of Parameter Search Spaces.}

As mentioned earlier, conventional ISPs are known to have many local
minima \cite{tseng2019hyperparameter}, and in fact, our work also
suffers from them. Specifically, as described in Appendix, completely
different ISP parameters were chosen when changing the initial random
seed, resulting in a slightly different accuracy. To solve this problem,
we propose an \emph{automatic tuning of parameter search spaces}.
The idea is to create search spaces that do not contain local minima
by making good use of the fact that our ISP functions are white-box,
unlike typical DNNs. Specifically, local minima are excluded by appropriately
narrowing the parameter search spaces $\left(P_{l,min},P_{l,max}\right)$.
The algorithm is shown in \cref{alg:tune}.

\begin{algorithm}[tb] 
\caption{Automatic Tuning of Parameter Search Spaces}\label{alg:tune} 
\SetAlgoLined
\PyCode{$P_{min}$,$P_{max}$ = init(initialize with enough large search spaces)} \\

\PyCode{for s in range(S)} \PyComment{S: total training stages} \\
	\Indp
	\PyCode{losses = array([T])} \\
    \PyCode{params = array([T,N,D])} \PyComment{N: \#ISP params, D: \#training data} \\
	\PyCode{for t in range(T)} \PyComment{train T times with different random seeds} \\
	\Indp
		\PyCode{l\_fs, p\_fs = train(model)} \\
		\PyCode{losses[t] = mean(l\_fs)} \PyComment{losses in the final epoch} \\
		\PyCode{params[t, :, :] = p\_fs} \PyComment{used ISP param. in the final epoch} \\
		\Indm
	\PyCode{$t_{best}$ = losses.argmin()} \\
    \PyComment{update search space} \\
	\PyCode{$P_{min}$ = r * params[$t_{best}$].min(dim=-1) + (1-r) * $P_{min}$} \\
    \PyCode{$P_{max}$ = r * params[$t_{best}$].max(dim=-1) + (1-r) * $P_{max}$} \\
\Indm
\end{algorithm}

In the first training stage, the search spaces of parameters are set
sufficiently wide, and the model is trained $T$ times with different
initial seeds. Then, we assume that the model with the best average
loss value in the final epoch is the model that falls into the best
local minima. We monitor the ISP parameters used by the best model
during the final epoch, and update the parameter search spaces. By
repeating this operation several times, search spaces excluding local
minima as much as possible is obtained. Our ISP has a small number
of parameters, so the loss value is not affected by overfitting, and
the algorithm works well. Although the algorithm needs multiple times
of training, the training cost is lower than other methods because
the model itself is lightweight and the convergence is quick as described
below.

\subsubsection*{Denoiser as Denoiser.}

We propose a training method for the case where DNN denoiser is used.
As shown in the experiments, in tasks where denoiser is not required,
the proposed method outperforms even large-scale DNN-based ISPs in
terms of accuracy. From this, it can be inferred that the DNN is inferior
to the proposed ISP control for color mapping. Therefore, when adding
a DNN as a denoiser, we make the DNN perform only denoising. First,
our model is trained with \emph{automatic tuning of parameter search
spaces} while removing the denoiser. It allows our model to learn
to map the color outside of the denoiser. Then, we add the denoiser
and freeze the non-denoiser parts. This allows the denoiser to learn
to perform only denoising. Finally, the whole model is finetuned.
The final finetuning improves the image quality a little because the
average luminance value can be changed due to noise removal and subsequent
non-linear processing.

\subsubsection*{Local L1 Loss.}

In addition to the aforementioned \emph{denoiser as denoiser}, the
following loss function is further added to restrict the denoiser
from changing color: 
\[
L_{LocalL1}(X)=L_{1}(AvgPool(I_{DN}(X)),AvgPool(X)),
\]
 where $AvgPool$ is the average pooling with kernel size $k_{a}$
and stride $s_{a}$.

\section{Experiments}

\subsection{Datasets}

The effectiveness of the proposed method is verified on various tasks
and datasets. We evaluate on four tasks: the newly defined \textbf{universal
ISP task}, where RAW images from any image sensors with unknown sensor
characteristics need to be processed; the \textbf{normal ISP task},
where RAW images from a specific sensor are processed; the \textbf{tone
mapping task}, where standardized CIE XYZ images after canceling sensor
characteristics and environment light are converted into sRGB; and
the \textbf{enhancement task}, where original sRGB images are converted
into high-quality sRGB images.

The universal ISP task is evaluated on the MIT-Adobe FiveK dataset
\cite{bychkovsky2011learning}. The dataset consists of RAW images
from 17 different image sensors paired with sRGB images from hardware
ISPs and manually retouched ground truth sRGB images. We follow the
convention on FiveK in the enhancement task \cite{chen2018deep,yang2022seplut,zeng2020learning};
the sRGB images retouched by expert C are used as the ground truth,
and 500 out of 5,000 images are defined as the test data. Since their
Bayer patterns are diverse per each sensor, we demosaic them as a
preprocessing and resize them isotropically into 480P size, whose
shorter edge length is 480. For the normal ISP task, the MAI21 dataset
\cite{ignatov2021learned} is used. In this dataset, the RAW images
from a smartphone sensor are inputs, and sRGB images from a high-quality
SLR camera are the ground truth. Since the MAI2022 challenge \cite{ignatov2022learned},
the last competition on the MA21 dataset, is finished and only the
training data is accessible, 10\% of the training data is defined
as the test data. As the images with close image IDs are patches cut
from the same image, the last 10\% of the image IDs are used as test
data without shuffling. For the tone mapping task, the FiveK and HDR+
\cite{hasinoff2016burst} datasets are used. HDR+ is a burst photography
dataset. Following previous works \cite{zeng2020learning,zhang2024lookup},
the intermediate 480P size 16-bit results after HDR composition are
used as the input, and JPEG images from a manually tuned HDR imaging
pipeline are used as the ground truth. The 247 out of 922 images are
defined as the test data. This dataset requires not only tone mapping
but also enhancement because some images are very dark. For the enhancement
task, the widely used FiveK dataset is used.

\subsection{Implementation Details}

\begin{table}[t]
\centering

\caption{Evaluation on FiveK dataset. The faster runtime on TensorRT or PyTorch
is reported. Improved runtimes with original CUDA kernels are written
in ($\:$). More details of runtime are reported in the Appendix.
The upper rows are general image restoration models for small images,
and the lower rows are specialized real-time models.}

\scalebox{0.91}{

\label{tab:fivek}

\setlength{\tabcolsep}{3pt}

\begin{tabular}{ccccccccc}
\hline 
 & \multicolumn{2}{c}{universal ISP} & \multicolumn{2}{c}{tone mapping} & \multicolumn{2}{c}{enhancement} & \multicolumn{2}{c}{runtime {[}ms{]}}\tabularnewline
 & PSNR & SSIM & PSNR & SSIM & PSNR & SSIM & 480P & 4K\tabularnewline
\hline 
\hline 
Restormer \cite{zamir2022restormer} &  &  &  &  & 24.13 &  &  & \tabularnewline
NAFNet-small \cite{chen2022simple} & 22.53 & 0.891 & 24.73 & 0.920 & 24.52 & 0.912 & 54.75 & 1428.60\tabularnewline
Retinexformer \cite{Cai_2023_ICCV} &  &  &  &  & 24.94 &  &  & \tabularnewline
\hline 
UPE \cite{wang2019underexposed} &  &  & 21.56 & 0.837 & 21.88 & 0.853 & 4.27 & 56.88\tabularnewline
MicroISP \cite{ignatov2022microisp} & 21.64 & 0.885 & 24.07 & 0.909 & 23.92 & 0.909 & 5.11 & 114.13\tabularnewline
HDRNet \cite{gharbi2017deep} &  &  & 24.52 & 0.915 & 24.66 & 0.915 & 3.49 & 56.07\tabularnewline
3D LUT \cite{zeng2020learning} &  &  & 25.06 & 0.920 & 25.21 & 0.922 & - (1.02) & - (1.14)\tabularnewline
CSRNet \cite{he2020conditional} &  &  & 25.19 & 0.921 & 25.17 & 0.921 & 3.09 & 77.10\tabularnewline
SYENet \cite{gou2023syenet} & 22.24 & 0.889 & 25.19 & 0.922 & 25.04 & 0.916 & 1.26 & 31.53\tabularnewline
RSFNet \cite{ouyang2023rsfnet} &  &  &  &  & 25.49 & 0.924 & 9.98 & \tabularnewline
AdaInt \cite{yang2022adaint} &  &  & 25.28 & 0.925 & 25.49 & 0.926 & - (1.29) & - (1.59)\tabularnewline
SepLUT \cite{yang2022seplut} & 22.82 & 0.891 & 25.43 & 0.922 & 25.47 & 0.921 & 6.18 (1.10) & 128.5 (1.20)\tabularnewline
F. Zhang \etal \cite{zhang2024lookup} &  &  & 25.53 & 0.907 &  &  &  & \tabularnewline
ours (global) & 22.99 & 0.901 & 25.35 & 0.928 & 24.80 & 0.918 & \textbf{1.00} & \textbf{4.49}\tabularnewline
ours (local) & \textbf{23.59} & \textbf{0.911} & \textbf{25.72} & \textbf{0.933} & \textbf{25.53} & \textbf{0.928} & 1.14 & 11.61\tabularnewline
\hline 
\end{tabular}}

\vspace{-4mm}
\end{table}

We use almost the same experimental settings for all tasks and datasets.
Our model is trained with the AdamW optimizer \cite{loshchilov2017decoupled}
from randomly initialized weights using a cosine annealing learning
rate schedule \cite{loshchilov2016sgdr}, whose maximum and minimum
learning rates are 1e-4 and 1e-7, with a linear warmup for the first
1,000 iterations. Random flip, rotation, and crop are applied to augment
the training data except for MAI21. On MAI21, no augmentations are
applied because the input is in Bayer pattern. Based on the dataset
sizes, we train 100 epochs on FiveK, 30 epochs on MAI21, and 600 epochs
on HDR+. The mini-batch size is 16. Although the 480P size of the
FiveK dataset contains various image sizes such as $480\times720$,
$720\times480$, $480\times640$, they are resized to $640\times480$
to achieve fast mini-batch training. During inference, original 480P
size images are used to evaluate. The denoiser is used only for MAI21,
a dataset that requires denoising. Its middle channel size is set
as 12. The loss function is a combination of mean square error loss
and perceptual loss \cite{johnson2016perceptual} with trained VGG16
\cite{simonyan2014very}: $L=L_{MSE}+0.1L_{VGG}$. When adding the
denoiser, the proposed \emph{local L1 loss} is added with a strength
of 0.01. For more details, please refer to the Appendix.

\subsection{Evaluation on FiveK Dataset}

\begin{figure}
\centering

\setlength{\tabcolsep}{0.5pt}

\begin{tabular}{ccccccc}
\includegraphics[width=0.13785\textwidth]{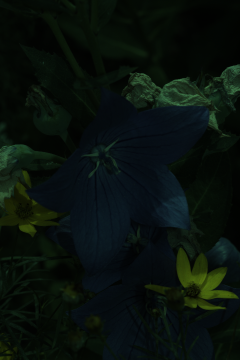} & \includegraphics[width=0.13785\textwidth]{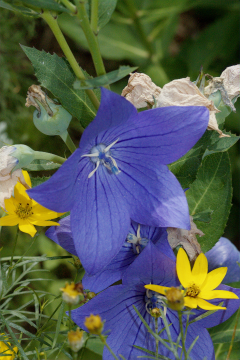} & \includegraphics[width=0.13785\textwidth]{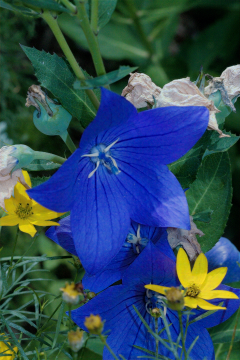} & \includegraphics[width=0.13785\textwidth]{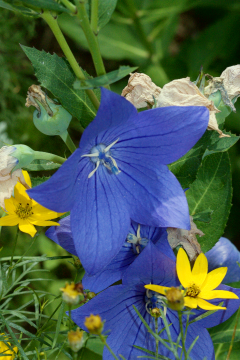} & \includegraphics[width=0.13785\textwidth]{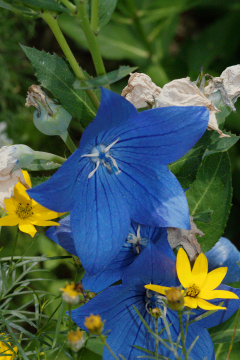} & \includegraphics[width=0.13785\textwidth]{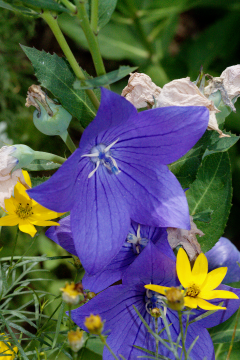} & \includegraphics[width=0.13785\textwidth]{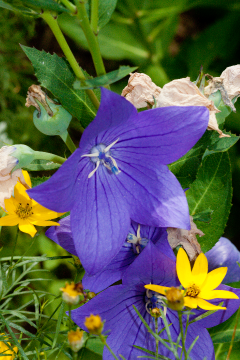}\tabularnewline
\includegraphics[width=0.13785\textwidth]{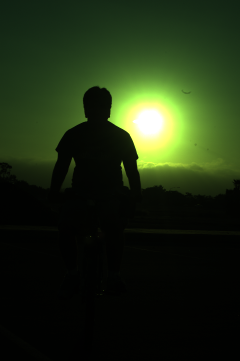} & \includegraphics[width=0.13785\textwidth]{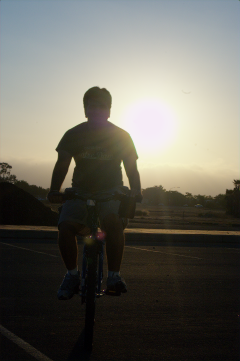} & \includegraphics[width=0.13785\textwidth]{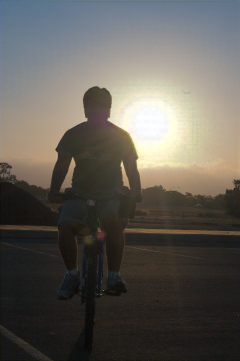} & \includegraphics[width=0.13785\textwidth]{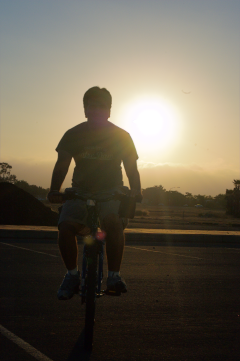} & \includegraphics[width=0.13785\textwidth]{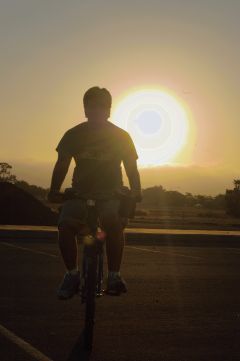} & \includegraphics[width=0.13785\textwidth]{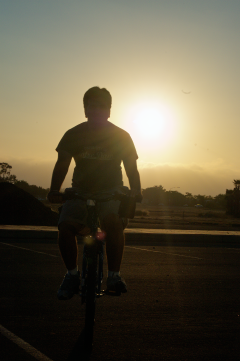} & \includegraphics[width=0.13785\textwidth]{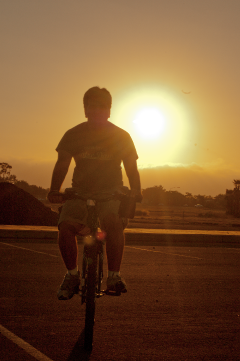}\tabularnewline
input & MicroISP & NAFNet & SYENet & SepLUT & ours & GT\tabularnewline
\end{tabular}

\vspace{-4mm}

\caption{Visualization of the univeral ISP task on FiveK \cite{bychkovsky2011learning}.
Our method estimates the true color well without artefacts. (Please
zoom in.)}

\label{fig:viz}
\end{figure}

The results on the three tasks are shown in Table \ref{tab:fivek}.
All tasks of NAFNet \cite{chen2022simple}, MicroISP \cite{ignatov2022microisp},
and SYENet \cite{yang2022seplut}, and the universal ISP task of SepLUT
\cite{yang2022seplut} are evaluated by us, and others are taken from
other papers. These models have slower convergence compared to ours,
so we trained them for 500 epochs following Retinexformer \cite{Cai_2023_ICCV}
to ensure fairness. The runtime is measured on a V100 GPU using two
implementations, Pytorch \cite{paszke2019pytorch} and TensorRT \cite{TensorRT},
and the faster one with float32 precision is described. For more details,
please refer to the Appendix.

Firstly, the universal ISP task is the most challenging, resulting
in low accuracy for all methods. However, our method significantly
outperforms existing methods, demonstrating its ability to cancel
sensor-specific characteristics through dynamic ISP control. DNN-based
ISPs with excessive degrees of freedom and SepLUT with more degrees
of freedom than our method do not learn well as shown Fig. \ref{fig:viz}.
Secondly, our method achieves the highest accuracy in the tone mapping
task among all tasks and methods. Since our method is designed as
an ISP, it performs exceptionally well in tone mapping, which is a
partial task of ISPs. Although our accuracy for the enhancement task
is not as good as that of the tone mapping task, it is more accurate
than existing methods. Further improvement can be made by enhancing
the inverse tone mapping function.

Despite achieving state-of-the-art accuracy, the runtime of our method
is significantly faster than existing methods. Look-up table-based
methods greatly improve runtime on high-resolution images by implementing
a CUDA kernel \cite{zeng2020learning,yang2022adaint,yang2022seplut,zhang2024lookup}.
Our method can also be improved by implementing a CUDA kernel. As
memory transfer costs are high for high-resolution images, combining
the entire ISP functions into a single CUDA kernel can lead to a significant
speed up. However, this is a trade-off with versatility, as devices
like smartphones do not support CUDA.

How the proposed components contribute to the performance is described
in Table \ref{tab:hdrplus}(a). \emph{Automatic tuning of parameter
search spaces} partially solves the local minima problem and improves
the accuracy. The proposed encoder also succeeds in generating better
features despite the size. Furthermore, the \emph{group-wise sigmoid
cross-attention} used as $f_{up}$ improves the accuracy over the
normal cross-attention as intended.

\begin{table}
\centering

\caption{(a) Ablation studies on FiveK and (b) the benchmark on HDR+ in the
tone mapping task. In the ablation study, the proposed \emph{automatic
tuning of parameter search spaces} (ATPS), encoder, and $f_{up}$
in the controller are evaluated. The more detailed ablation studies
are reported in the Appendix.}

\scalebox{0.92}{

\label{tab:hdrplus}

\setlength{\tabcolsep}{2pt}

\begin{tabular}{cc}
(a) ablation studies on FiveK & (b) benchmark on HDR+\tabularnewline
\begin{tabular}{ccccc}
\hline 
ATPS & %
\begin{tabular}{c}
encoder\tabularnewline
block\tabularnewline
\end{tabular} & $f_{up}$ & PSNR & %
\begin{tabular}{c}
runtime\tabularnewline
4K\tabularnewline
\end{tabular}\tabularnewline
\hline 
\hline 
 & $\checkmark$ & $\checkmark$ & 25.62 & 11.61\tabularnewline
$\checkmark$ & \cite{chen2022simple} & $\checkmark$ & 25.30 & 12.28\tabularnewline
$\checkmark$ & \cite{gou2023syenet} & $\checkmark$ & NaN & \tabularnewline
$\checkmark$ & $\checkmark$ & cross-attn & 25.64 & 11.65\tabularnewline
$\checkmark$ & $\checkmark$ & \cite{yoshimura2023dynamicisp} & 25.63 & 11.59\tabularnewline
$\checkmark$ & $\checkmark$ & $\checkmark$ & 25.72 & 11.61\tabularnewline
\hline 
\end{tabular} & %
\begin{tabular}{ccc}
\hline 
 & PSNR & SSIM\tabularnewline
\hline 
\hline 
CSRNet \cite{he2020conditional} & 23.72 & 0.864\tabularnewline
3D LUT \cite{zeng2020learning} & 23.54 & 0.885\tabularnewline
HDRNet \cite{gharbi2017deep} & 24.14 & 0.845\tabularnewline
DeepLPF \cite{moran2020deeplpf} & 25.73 & 0.902\tabularnewline
CLUT-Net \cite{zhang2022clut} & 26.05 & 0.892\tabularnewline
F. Zhang \etal \cite{zhang2024lookup} & 26.62 & 0.907\tabularnewline
ours (global) & 22.26 & 0.869\tabularnewline
ours (local) & \textbf{26.72} & \textbf{0.927}\tabularnewline
\hline 
\end{tabular}\tabularnewline
\end{tabular}}

\vspace{-3mm}
\end{table}

\subsection{Evaluation of the Tone Mapping Task Including Low-light Enhancement
on HDR+ Dataset}

Since HDR+ contains more HDR and darker scenes than the FiveK tone
mapping task, the ISP parameters need to be controlled more dynamically.
Our method succeeds in controlling them appropriately and achieves
state-of-the-art accuracy, as shown in Table \ref{tab:hdrplus}(b).
There is a significant advantage of local control over global control
in HDR+ compared to FiveK. This is probably because the input and
ground truth of HDR+ are created from different source data, \ie
several burst RAW images are combined with different proportions,
which requires non-global mapping. On the other hand, the ground truth
of FiveK is created by hand-tuning global parameters, so in principle,
global control alone is sufficient. However, the degree of freedom
of the ISP function is limited, so local control may have improved
the accuracy somewhat. To maximize the benefits of the proposed local
control, it is necessary to prepare datasets like HDR+ or more locally
tuned datasets.

\subsection{Evaluation of the Normal ISP Task on MAI21 Dataset}

On MAI21, our ISP including the proposed denoiser is evaluated. MicroISP
and SYENet are trained for 200 epochs instead of 30 epochs, following
the original setup of SYNet. As shown in Table \ref{tab:mai21}(a),
our work also outperforms in tasks that require a denoiser. Unlike
the results on FiveK, our method with local control is slower than
SYENet. This is because our method performs the demosaicing included
in our denoiser first, and the high-resolution image has to be processed
in the remaining ISP functions. It can be solved if the demosaicing
is done at the end, but it will become a trade-off between accuracy
and speed. Furthermore, as noted in the appendix, the number of operations
in our ISP is significantly less than that of CNNs and look-up tables
and is only slow because small calculations are spread out. If an
optimized implementation per backends such as combining ISP functions
into a single CUDA, similar to 3D LUT \cite{zeng2020learning} and
SepLUT\cite{yang2022seplut}, the speed would increase dramatically.

Next, we discuss the additional DNN denoiser in our ISP. First, as
shown in \ref{tab:mai21}(b), controlling the dynamic filter with
the proposed local control significantly improves the accuracy. Although
it increases the runtime, the accuracy is better than using SYENet
as the denoiser of our ISP. Furthermore, the increased runtime is
due to our implementation using an unfold function and matrix multiplication,
which is convertible to various backends. Our dynamic filter needs
less computation than DDF \cite{zhou2021decoupled}, which is faster
than standard convolution with an optimal implementation. Next, the
accuracy deteriorates when the denoiser is trained at the same time.
This is because the DNN denoiser performs part of the tone mapping,
even though our simple ISP functions are superior to DNNs in terms
of tone mapping, as shown in the evaluation on FiveK. The proposed
\emph{denoiser as denoiser} and \emph{local L1 loss} improve the accuracy
by letting the DNN denoiser only denoise.

\begin{table}
\centering

\caption{Evaluation on MAI21 in the normal ISP task. (a) The top four methods
in the MAI2022 ISP challenge report \cite{ignatov2022learned} are
listed. (b) The ablation studies regarding the denoiser in our ISP
are conducted, including the proposed training methods of \emph{denoiser
as denoiser} (DasD) and \emph{local L1 loss} (LocalL1).}

\scalebox{0.81}{

\setlength{\tabcolsep}{1.5pt}

\label{tab:mai21}

\begin{tabular}{cc}
(a) benchmark & (b) ablation studies\tabularnewline
\begin{tabular}{ccccccc}
\hline 
 & \multicolumn{2}{c}{official test} & \multicolumn{2}{c}{our split} & \multicolumn{2}{c}{runtime (full HD)}\tabularnewline
 & score & PSNR & PSNR & SSIM & mobile & V100\tabularnewline
\hline 
\hline 
MicroISP \cite{ignatov2022microisp} & 9.25 & 23.87 & 23.00 & 0.799 & 23.1 & 8.28\tabularnewline
ENERZAi & 10.27 & 23.80 &  &  & 18.9 & \tabularnewline
MiAlgo & 14.87 & 23.30 &  &  & 6.8 & \tabularnewline
SYENet \cite{gou2023syenet} & 21.24 & 23.96 & 24.32 & 0.860 & 11.4 & \textbf{2.39}\tabularnewline
ours (global) &  &  & 23.67 & 0.835 &  & 2.76\tabularnewline
ours (local) &  &  & \textbf{24.63} & \textbf{0.864} &  & 6.64\tabularnewline
\hline 
\end{tabular} & %
\begin{tabular}{ccccc}
\hline 
type & DasD & LocalL1 & PSNR & runtime\tabularnewline
\hline 
\hline 
none &  &  & 24.18 & 3.26\tabularnewline
SYENet \cite{gou2023syenet} &  &  & 24.08 & 6.03\tabularnewline
SYENet \cite{gou2023syenet} & $\checkmark$ & $\checkmark$ & 24.59 & 6.03\tabularnewline
wo dyFilter & $\checkmark$ & $\checkmark$ & 24.47 & 4.38\tabularnewline
w/ dyFilter &  &  & 24.07 & 6.64\tabularnewline
w/ dyFilter & $\checkmark$ &  & 24.61 & 6.64\tabularnewline
w/ dyFilter & $\checkmark$ & $\checkmark$ & 24.63 & 6.64\tabularnewline
\hline 
\end{tabular}\tabularnewline
\end{tabular}}

\vspace{-6mm}
\end{table}

\section{Conclusion}

This study proposes a new type of ISP for perceptual image quality
that consists of very simple ISP functions, but the parameters are
dynamically controlled to achieve high image quality. Unlike ISPs
for image recognition, ISPs for perceptual quality require a great
deal of processing to be done accurately. Our work achieves this by
combining the proposed ISP functions, encoder, controller, and training
method. Additionally, although not sophisticated, our work also performs
well in the enhancement task by adding an inverse tone mapping function,
demonstrating its versatility. Our work, which combines state-of-the-art
lightness and state-of-the-art accuracy, will be useful in various
situations, from applications in edge devices to operating in the
cloud. Furthermore, the fact that it outperforms large-scale DNNs
in terms of accuracy raises questions about the color-mapping capabilities
of DNNs and may contribute to the future improvement of DNNs.

\appendix

\chapter*{Appendix of \textquotedblleft PQDynamicISP: Dynamically Controlled
Image Signal Processor for Any Image Sensors Pursuing Perceptual Quality\textquotedblright}

\section{More Implementation Details}

Most of the implementation details are written in Section 4.2. More
implementation details are given below.

The output channel size of each block of the encoder is set to 24,
48, and 96. The latent variable $V_{l}$ in the controller is set
to 256 for global control, as it is computationally inexpensive, and
to 96 for local control. The kernel size $k_{a}$ and stride $s_{a}$
of the \emph{local L1 loss} is set to 16 and 8. In the \emph{automatic
tuning of parameter search spaces,} two stages of training are performed,
with five and four scratch training in each stage. Since the sigmoid
function used for constraining search spaces in eq. 5 vanishes the
gradient of the edge of the domain, it was observed that the accuracy
deteriorates if the search spaces are too narrow. So, we set $r=0.7$
in the Algorithm 2.

PSNR and SSIM are used as metrics. PSNR is calculated at the original
resolution for all the datasets. On the other hand, SSIM is calculated
differently per dataset following previous works. For the FiveK dataset,
we calculate the SSIM after applying a low-pass filter and subsampling
to $256\times256$ size following previous works \cite{yang2022seplut,yang2022adaint,zeng2020learning}.
For the HDR+ dataset, our table is mainly based on F. Zhang \etal
\cite{zhang2024lookup}, and they seem to calculate the SSIM at the
original resolution. So we calculate it at the original resolution.
For the MAI21 dataset, we calculate it at the original resolution
because the original images are $256\times256$ patches.

\section{Ablation Studies on Local Control}

The optimal resolution of the local control is investigated in addition
to the effectiveness of adding decoder blocks in the encoder. We use
a pixel shuffle \cite{shi2016real} in our decoder block to upsample.
As shown in Table \ref{tab:local}, local control at a resolution
of $28\times28$ is optimal, and when we try to control more finely,
it becomes unstable and the accuracy deteriorates. It can also be
seen that the accuracy is higher when no decoder blocks are used.
This may be because the global features are well obtained by the attention
in the encoder blocks, and the accuracy deteriorates as more global
information is added by the decoder blocks. Although it could be improved
by introducing skipping connections similar to Unet \cite{ronneberger2015u},
we only use encoder blocks to save computational cost.

\begin{table}[t]
\centering

\caption{Ablation studies on the optimal resolution of the local control and
additional decoder blocks. The numbers in {[}$\;${]} means the number
of blocks in each encoder/decoder stage. The ablation study is conducted
for the tone mapping task on FiveK dataset \cite{bychkovsky2011learning}.}

\scalebox{1.0}{

\label{tab:local}

\setlength{\tabcolsep}{3pt}

\begin{tabular}{cccc}
\hline 
\multicolumn{2}{c}{encoder architecture} & \multirow{2}{*}{%
\begin{tabular}{c}
controller's\tabularnewline
resolution\tabularnewline
\end{tabular}} & \multirow{2}{*}{PSNR}\tabularnewline
encoder blocks & decoder blocks &  & \tabularnewline
\hline 
\hline 
{[}1, 1{]} &  & 56$\times$56 & 25.60\tabularnewline
{[}1, 1, 1, 1{]} & {[}1, 1{]} & 56$\times$56 & 25.47\tabularnewline
{[}1, 1, 1{]} &  & 28$\times$28 & \textbf{25.72}\tabularnewline
{[}1, 1, 1, 1{]} & {[}1{]} & 28$\times$28 & 25.57\tabularnewline
{[}1, 1, 1, 1{]} &  & 14$\times$14 & 25.44\tabularnewline
\hline 
\end{tabular}}

\vspace{-1mm}
\end{table}

\section{Local Minima in the ISP}

\begin{figure}
\centering

\includegraphics[width=1\textwidth]{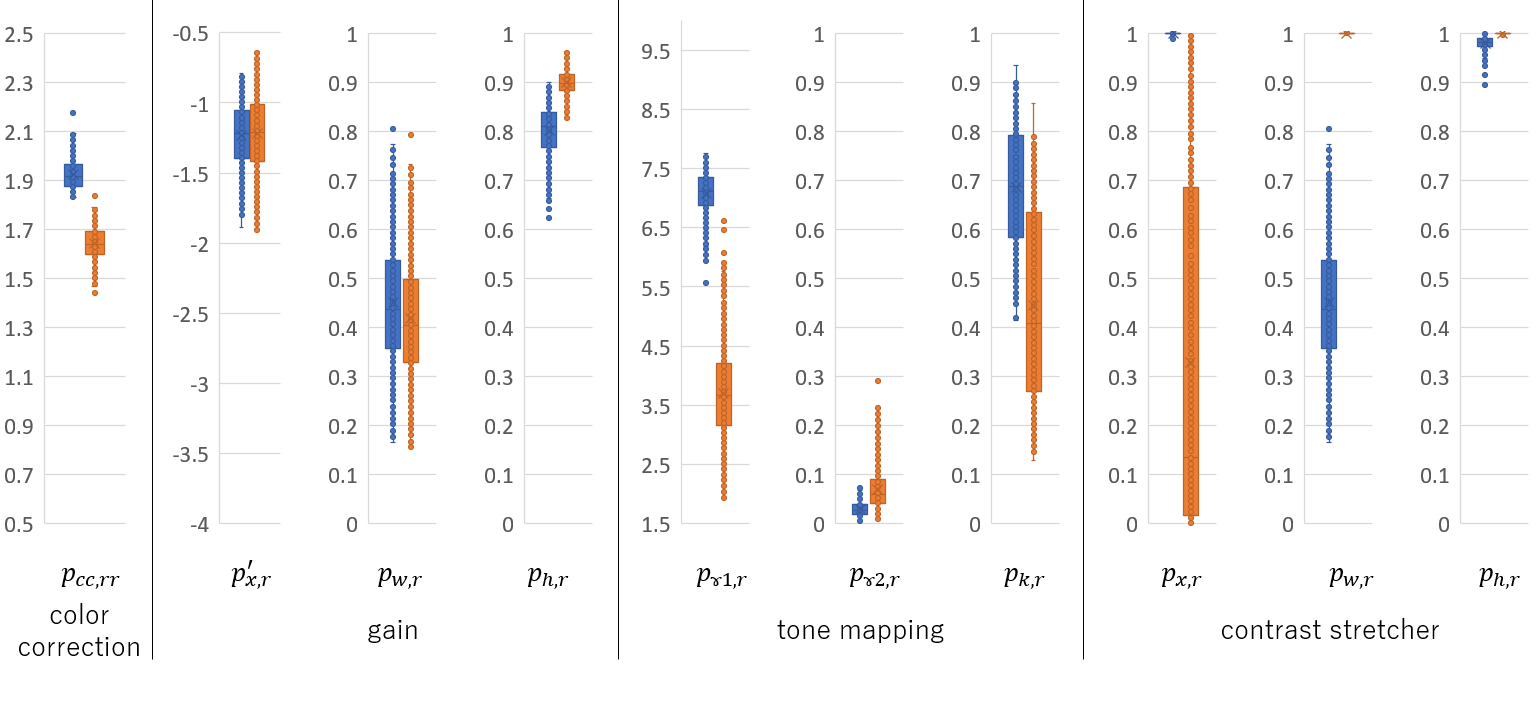}

\vspace{-1mm}

\caption{For part of ISP parameters, the distributions of used parameters during
test time are shown. The parameters used by two models with similar
accuracy are plotted as blue (PSNR 26.2) and orange (PSNR 26.4). The
existence of local minima is supported by the fact that they use different
parameters but with almost identical accuracy.}

\label{fig:usedparam}
\end{figure}

In this section, how much the ISP parameters are controlled and how
local minima exist in the ISP are discussed using Fig. \ref{fig:usedparam}.
It shows the distributions of ISP parameters used by two models trained
with different seeding. First, both models successfully control parameters,
and that's why our model achieves high accuracy with simple ISP functions.
Second, although the accuracies of the two models are similar, the
used ISP parameters are different. Similar accuracies with different
parameters indicate that there are local minima. Furthermore, you
can see that some local minima are quite far from each other which
can not be overcome by devising an optimizer. So, it is preferable
to train multiple times and choose a model that reaches better local
minima by using \emph{automatic tuning of parameter search}.

\section{Details of Runtime}

The faster runtime on TensorRT \cite{TensorRT} or PyTorch \cite{paszke2019pytorch}
is reported in Table 1 of the main paper. Here, we report the details.
As shown in Table \ref{tab:runtime}, by converting from PyTorch models
to TensorRT models, a significant speed improvement is observed for
our model over other models. There are two possible reasons for this.
One is that although the FLOPS of our controller is small thanks to
the very small resolution, the number of functions in the controller
is large. Pytorch implementation calls a lot of Python interfaces,
which are time-consuming. By converting to TensorRT, the cost is reduced.
In addition, in the ISP functions, although the FLOPS is less than
one layer of convolution, simple operations such as quadrature operations
are used a lot. In this case, because the images are high-resolution
images, the memory transfer costs between functions are considered
to be the reason for the slowdown. The speed could be further improved
by combining the ISPs into a single kernel as described in the main
paper. In fact, SepLUT \cite{yang2022seplut} improves the speed from
51.19 ms to 1.20 ms by creating a single CUDA kernel. While SepLUT
uses a 1D look-up table with 17 grids and a 3D look-up table with
$17\times17\times3$ grids, our gain and contrast stretcher can be
represented by a 1D look-up table with four grids.

\begin{table}[t]
\centering

\caption{The detail of the runtime on two backends.}

\scalebox{0.91}{

\label{tab:runtime}

\setlength{\tabcolsep}{3pt}

\begin{tabular}{ccccc}
\hline 
 & \multicolumn{2}{c}{Pytorch runtime {[}ms{]}} & \multicolumn{2}{c}{TensorRT runtime {[}ms{]}}\tabularnewline
 & 480P & 4K & 480P & 4K\tabularnewline
\hline 
NAFNet-small \cite{chen2022simple} & 54.86 & 1428.60 & 54.75 & OOM\tabularnewline
MicroISP \cite{ignatov2022microisp} & 12.89 & 270.30 & 5.11 & 114.13\tabularnewline
SYENet \cite{gou2023syenet} & 6.85 & 166.70 & 1.26 & 31.53\tabularnewline
SepLUT \cite{yang2022seplut} & \textbf{6.18} & 128.5 & 31.1 & 804.60\tabularnewline
ours (global) & 6.98 & \textbf{107.9} & \textbf{1.00} & \textbf{4.49}\tabularnewline
ours (local) & 11.20 & 172.00 & 1.14 & 11.61\tabularnewline
\hline 
\end{tabular}}

\vspace{-1mm}
\end{table}

\section{More Qualitative Comparison}

\begin{figure}
\centering

\setlength{\tabcolsep}{4pt}

\renewcommand{\arraystretch}{0.3}

\scalebox{1.09}{

\begin{tabular}{cccc}
\begin{tabular}{c}
input\tabularnewline
\end{tabular} & \includegraphics[height=0.143\textheight]{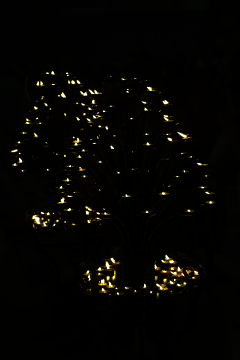} & \includegraphics[height=0.143\textheight]{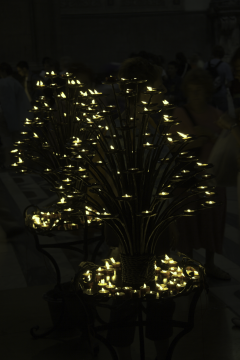} & \includegraphics[height=0.143\textheight]{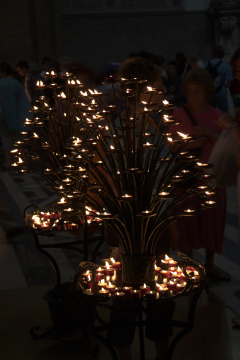}\tabularnewline
\begin{tabular}{c}
NAFNet\tabularnewline
\end{tabular} & \includegraphics[height=0.143\textheight]{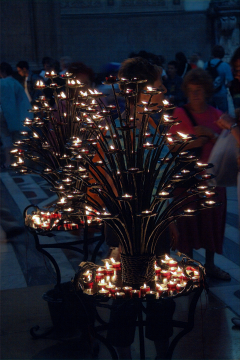} & \includegraphics[height=0.143\textheight]{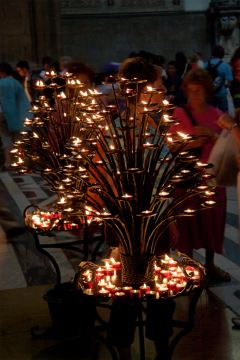} & \includegraphics[height=0.143\textheight]{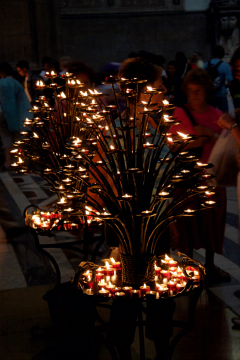}\tabularnewline
\begin{tabular}{c}
SYENet\tabularnewline
\end{tabular} & \includegraphics[height=0.143\textheight]{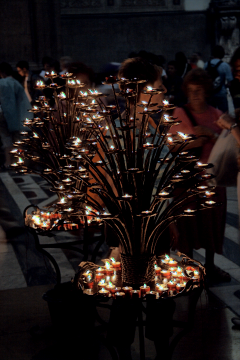} & \includegraphics[height=0.143\textheight]{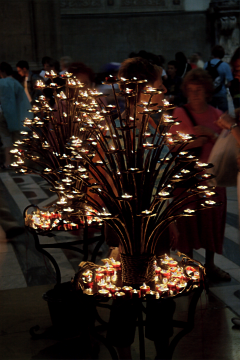} & \includegraphics[height=0.143\textheight]{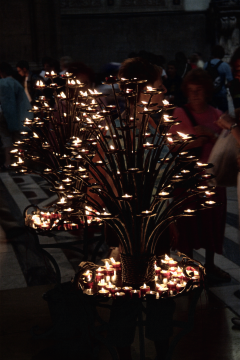}\tabularnewline
\begin{tabular}{c}
SepLUT\tabularnewline
\end{tabular} & \includegraphics[height=0.143\textheight]{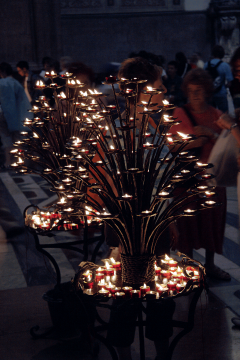} & \includegraphics[height=0.143\textheight]{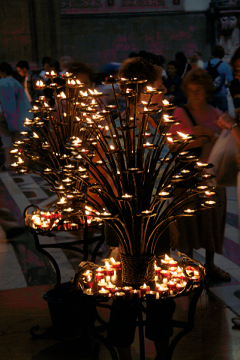} & \includegraphics[height=0.143\textheight]{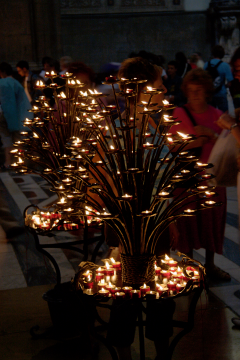}\tabularnewline
\begin{tabular}{c}
ours\tabularnewline
\end{tabular} & \includegraphics[height=0.143\textheight]{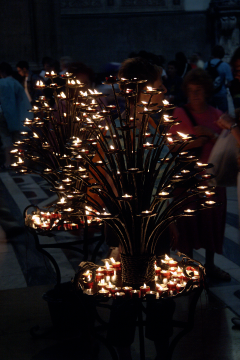} & \includegraphics[height=0.143\textheight]{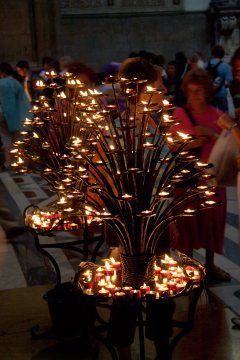} & \includegraphics[height=0.143\textheight]{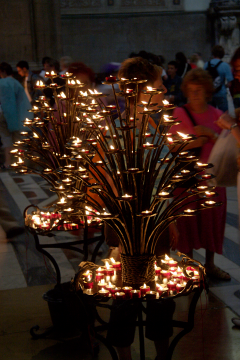}\tabularnewline
\begin{tabular}{c}
GT\tabularnewline
\end{tabular} & \includegraphics[height=0.143\textheight]{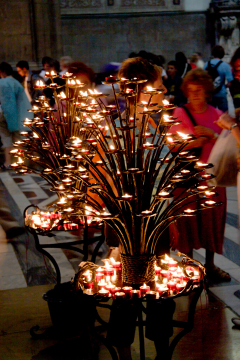} & \includegraphics[height=0.143\textheight]{figs/viz_fivek/a0498_gt_srgb} & \includegraphics[height=0.143\textheight]{figs/viz_fivek/a0498_gt_srgb}\tabularnewline
 & general ISP & tone mapping & enhancement\tabularnewline
\end{tabular}}

\vspace{-1mm}

\caption{Visual comparisons on FiveK \cite{bychkovsky2011learning}.}

\label{fig:viz-1}
\end{figure}

\begin{figure}
\centering

\renewcommand{\arraystretch}{0.9}

\setlength{\tabcolsep}{0.5pt}

\begin{tabular}{ccccc}
\includegraphics[width=0.198\textwidth]{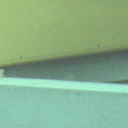} & \includegraphics[width=0.198\textwidth]{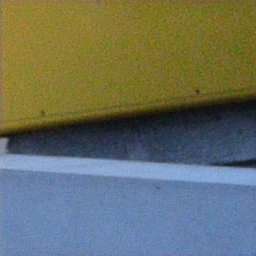} & \includegraphics[width=0.198\textwidth]{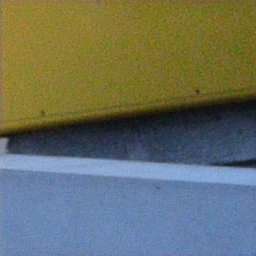} & \includegraphics[width=0.198\textwidth]{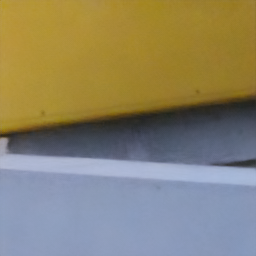} & \includegraphics[width=0.198\textwidth]{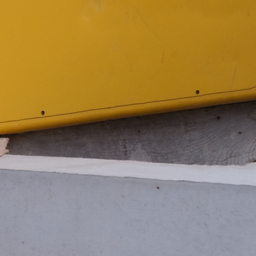}\tabularnewline
\includegraphics[width=0.198\textwidth]{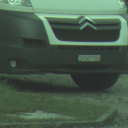} & \includegraphics[width=0.198\textwidth]{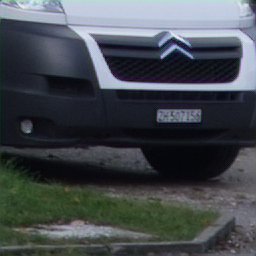} & \includegraphics[width=0.198\textwidth]{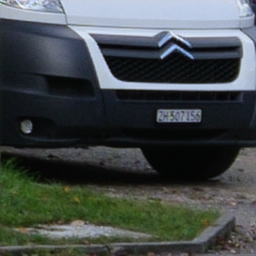} & \includegraphics[width=0.198\textwidth]{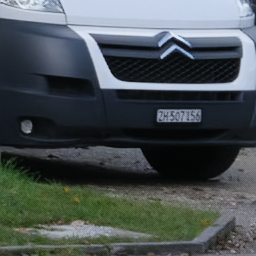} & \includegraphics[width=0.198\textwidth]{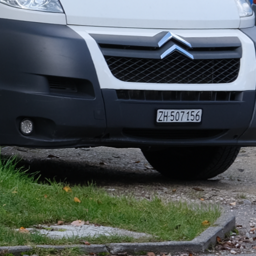}\tabularnewline
\includegraphics[width=0.198\textwidth]{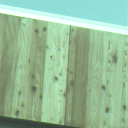} & \includegraphics[width=0.198\textwidth]{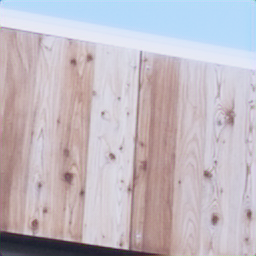} & \includegraphics[width=0.198\textwidth]{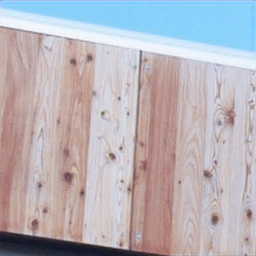} & \includegraphics[width=0.198\textwidth]{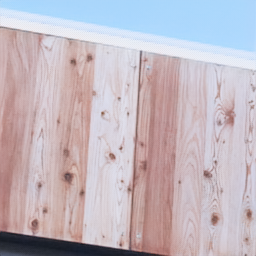} & \includegraphics[width=0.198\textwidth]{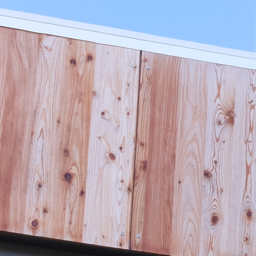}\tabularnewline
\includegraphics[width=0.198\textwidth]{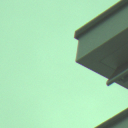} & \includegraphics[width=0.198\textwidth]{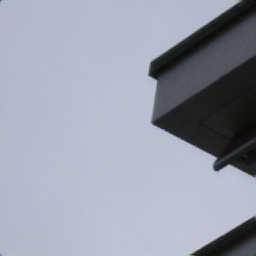} & \includegraphics[width=0.198\textwidth]{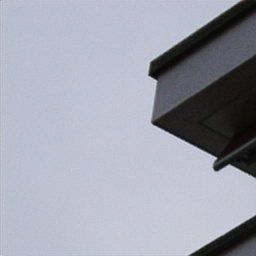} & \includegraphics[width=0.198\textwidth]{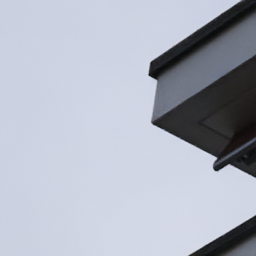} & \includegraphics[width=0.198\textwidth]{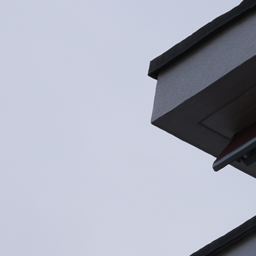}\tabularnewline
input & MicroISP & SYENet & ours & GT\tabularnewline
\end{tabular}

\vspace{-1mm}

\caption{Visual comparisons on MAI21 dataset \cite{ignatov2021learned}. The
input RAW images are from a smartphone sensor, and the ground truth
sRGB images are from a high-end DSLR camera. For visualization, we
apply gain and gamma on the input RAW images.}

\label{fig:viz-1-2}
\end{figure}

\begin{figure}
\centering

\renewcommand{\arraystretch}{0.9}

\setlength{\tabcolsep}{0.5pt}

\begin{tabular}{cccc}
\includegraphics[width=0.248\textwidth]{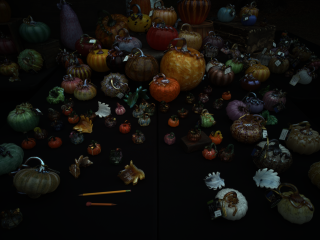} & \includegraphics[width=0.248\textwidth]{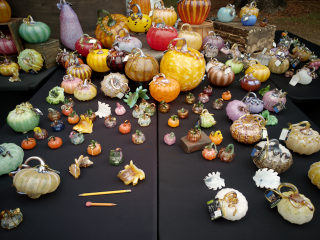} & \includegraphics[width=0.248\textwidth]{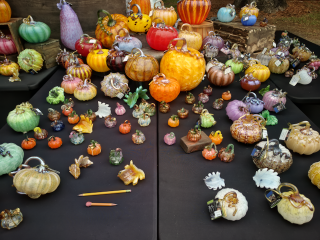} & \includegraphics[width=0.248\textwidth]{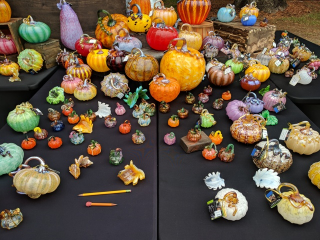}\tabularnewline
\includegraphics[width=0.248\textwidth]{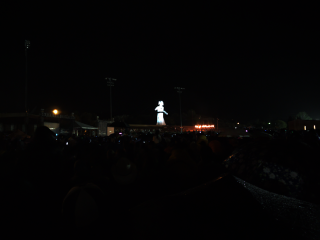} & \includegraphics[width=0.248\textwidth]{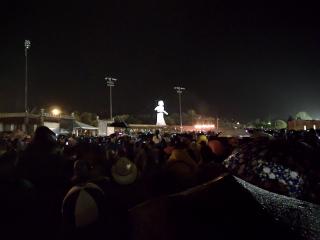} & \includegraphics[width=0.248\textwidth]{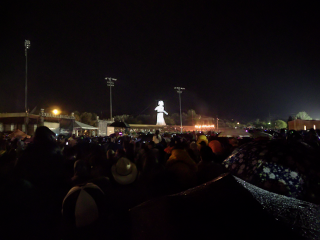} & \includegraphics[width=0.248\textwidth]{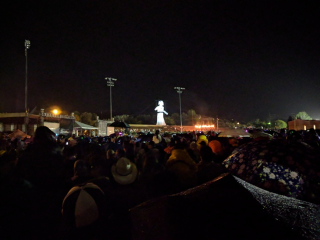}\tabularnewline
\includegraphics[width=0.248\textwidth]{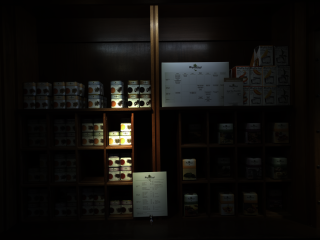} & \includegraphics[width=0.248\textwidth]{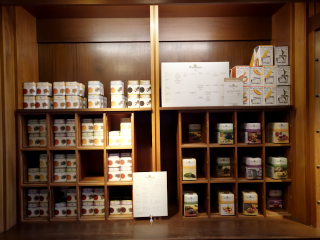} & \includegraphics[width=0.248\textwidth]{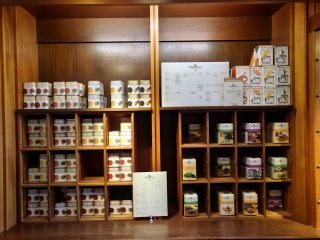} & \includegraphics[width=0.248\textwidth]{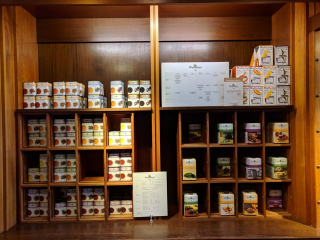}\tabularnewline
\includegraphics[width=0.248\textwidth]{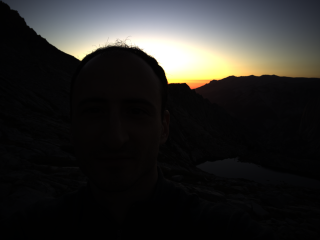} & \includegraphics[width=0.248\textwidth]{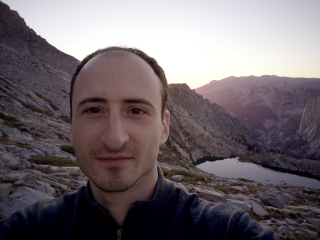} & \includegraphics[width=0.248\textwidth]{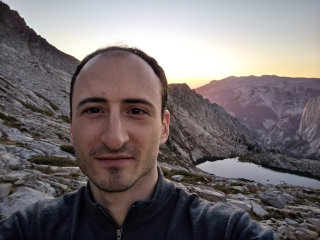} & \includegraphics[width=0.248\textwidth]{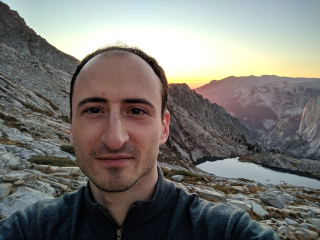}\tabularnewline
input & ours (global) & ours (local) & GT\tabularnewline
\end{tabular}

\vspace{-1mm}

\caption{Visual comparisons of our local and global control on HDR+ dataset
\cite{hasinoff2016burst}. We sellect bright, dark, and two HDR scenes.
In HDR scenes, our local control works better.}

\label{fig:viz-1-2-1}
\end{figure}

Visual comparisons on the universal ISP task, tone mapping task, and
enhancement task with FiveK dataset \cite{bychkovsky2011learning}
are shown in Fig. \ref{fig:viz-1}. Also, the normal ISP task on MAI21
dataset and the tone mapping task on HDR+ dataset are shown in Fig.
\ref{fig:viz-1-2} and Fig. \ref{fig:viz-1-2-1}. Fig. \ref{fig:viz-1-2}
shows that our denoiser is working properly, not only denoising but
also working for higher resolution. Although the accuracy of the denoiser
looks not as good as that of recent large-scale DNNs, it works well
considering the light weight. Fig. \ref{fig:viz-1-2} shows the difference
between the proposed global and local control. It can be seen that
the local control is able to achieve vivid colors in high dynamic
range (HDR) scenes, although whether the environment is bright or
dark does not make much difference in accuracy.

\clearpage

\clearpage

\bibliographystyle{splncs04}
\bibliography{main}

\begin{thebibliography}{10}
\providecommand{\url}[1]{\texttt{#1}}
\providecommand{\urlprefix}{URL }
\providecommand{\doi}[1]{https://doi.org/#1}

\bibitem{TensorRT}
{NVIDIA TensorRT}. \url{https://developer.nvidia.com/tensorrt}, [Online;
  accessed 10-February-2024]

\bibitem{onnxruntime}
{ONNX Runtime}. \url{https://onnxruntime.ai/}, [Online; accessed
  10-February-2024]

\bibitem{tflite}
{TensorFlow Lite}. \url{https://www.tensorflow.org/lite}, [Online; accessed
  10-February-2024]

\bibitem{afifi2021cross}
Afifi, M., Barron, J.T., LeGendre, C., Tsai, Y.T., Bleibel, F.: Cross-camera
  convolutional color constancy. In: Proceedings of the IEEE/CVF International
  Conference on Computer Vision. pp. 1981--1990 (2021)

\bibitem{buades2005non}
Buades, A., Coll, B., Morel, J.M.: A non-local algorithm for image denoising.
  In: 2005 IEEE computer society conference on computer vision and pattern
  recognition (CVPR'05). vol.~2, pp. 60--65. Ieee (2005)

\bibitem{bychkovsky2011learning}
Bychkovsky, V., Paris, S., Chan, E., Durand, F.: Learning photographic global
  tonal adjustment with a database of input/output image pairs. In: CVPR 2011.
  pp. 97--104. IEEE (2011)

\bibitem{Cai_2023_ICCV}
Cai, Y., Bian, H., Lin, J., Wang, H., Timofte, R., Zhang, Y.: Retinexformer:
  One-stage retinex-based transformer for low-light image enhancement. In:
  Proceedings of the IEEE/CVF International Conference on Computer Vision
  (ICCV). pp. 12504--12513 (October 2023)

\bibitem{chen2022simple}
Chen, L., Chu, X., Zhang, X., Sun, J.: Simple baselines for image restoration.
  In: European Conference on Computer Vision. pp. 17--33. Springer (2022)

\bibitem{chen2018deep}
Chen, Y.S., Wang, Y.C., Kao, M.H., Chuang, Y.Y.: Deep photo enhancer: Unpaired
  learning for image enhancement from photographs with gans. In: Proceedings of
  the IEEE conference on computer vision and pattern recognition. pp.
  6306--6314 (2018)

\bibitem{cheng2014illuminant}
Cheng, D., Prasad, D.K., Brown, M.S.: Illuminant estimation for color
  constancy: why spatial-domain methods work and the role of the color
  distribution. JOSA A  \textbf{31}(5),  1049--1058 (2014)

\bibitem{conde2022model}
Conde, M.V., McDonagh, S., Maggioni, M., Leonardis, A., P{\'e}rez-Pellitero,
  E.: Model-based image signal processors via learnable dictionaries. In:
  Proceedings of the AAAI Conference on Artificial Intelligence. vol.~36, pp.
  481--489 (2022)

\bibitem{conde2023nilut}
Conde, M.V., Vazquez-Corral, J., Brown, M.S., Timofte, R.: Nilut: Conditional
  neural implicit 3d lookup tables for image enhancement. arXiv preprint
  arXiv:2306.11920  (2023)

\bibitem{dabov2007image}
Dabov, K., Foi, A., Katkovnik, V., Egiazarian, K.: Image denoising by sparse
  3-d transform-domain collaborative filtering. IEEE Transactions on image
  processing  \textbf{16}(8),  2080--2095 (2007)

\bibitem{dai2020awnet}
Dai, L., Liu, X., Li, C., Chen, J.: Awnet: Attentive wavelet network for image
  isp. In: Computer Vision--ECCV 2020 Workshops: Glasgow, UK, August 23--28,
  2020, Proceedings, Part III 16. pp. 185--201. Springer (2020)

\bibitem{dinh2023end}
Dinh, K.Q., Choi, K.P.: End-to-end single-frame image signal processing for
  high dynamic range scenes. In: Proceedings of the IEEE/CVF Winter Conference
  on Applications of Computer Vision. pp. 2449--2458 (2023)

\bibitem{drago2003adaptive}
Drago, F., Myszkowski, K., Annen, T., Chiba, N.: Adaptive logarithmic mapping
  for displaying high contrast scenes. In: Computer graphics forum. vol.~22,
  pp. 419--426. Wiley Online Library (2003)

\bibitem{ershov2022ntire}
Ershov, E., Savchik, A., Shepelev, D., Bani{\'c}, N., Brown, M.S., Timofte, R.,
  Ko{\v{s}}{\v{c}}evi{\'c}, K., Freeman, M., Tesalin, V., Bocharov, D., et~al.:
  Ntire 2022 challenge on night photography rendering. In: Proceedings of the
  IEEE/CVF Conference on Computer Vision and Pattern Recognition. pp.
  1287--1300 (2022)

\bibitem{ershov2023physically}
Ershov, E., Tesalin, V., Ermakov, I., Brown, M.S.: Physically-plausible
  illumination distribution estimation. In: Proceedings of the IEEE/CVF
  International Conference on Computer Vision. pp. 12928--12936 (2023)

\bibitem{gharbi2017deep}
Gharbi, M., Chen, J., Barron, J.T., Hasinoff, S.W., Durand, F.: Deep bilateral
  learning for real-time image enhancement. ACM Transactions on Graphics (TOG)
  \textbf{36}(4),  1--12 (2017)

\bibitem{gou2023syenet}
Gou, W., Yi, Z., Xiang, Y., Li, S., Liu, Z., Kong, D., Xu, K.: Syenet: A simple
  yet effective network for multiple low-level vision tasks with real-time
  performance on mobile device. In: Proceedings of the IEEE/CVF International
  Conference on Computer Vision. pp. 12182--12195 (2023)

\bibitem{gu2012local}
Gu, B., Li, W., Zhu, M., Wang, M.: Local edge-preserving multiscale
  decomposition for high dynamic range image tone mapping. IEEE Transactions on
  image Processing  \textbf{22}(1),  70--79 (2012)

\bibitem{hansen2021isp4ml}
Hansen, P., Vilkin, A., Krustalev, Y., Imber, J., Talagala, D., Hanwell, D.,
  Mattina, M., Whatmough, P.N.: Isp4ml: The role of image signal processing in
  efficient deep learning vision systems. In: 2020 25th International
  Conference on Pattern Recognition (ICPR). pp. 2438--2445. IEEE (2021)

\bibitem{hasinoff2016burst}
Hasinoff, S.W., Sharlet, D., Geiss, R., Adams, A., Barron, J.T., Kainz, F.,
  Chen, J., Levoy, M.: Burst photography for high dynamic range and low-light
  imaging on mobile cameras. ACM Transactions on Graphics (Proc. SIGGRAPH Asia)
   \textbf{35}(6) (2016)

\bibitem{he2020conditional}
He, J., Liu, Y., Qiao, Y., Dong, C.: Conditional sequential modulation for
  efficient global image retouching. In: Computer Vision--ECCV 2020: 16th
  European Conference, Glasgow, UK, August 23--28, 2020, Proceedings, Part XIII
  16. pp. 679--695. Springer (2020)

\bibitem{hevia2020optimization}
Hevia, L.V., Patricio, M.A., Molina, J.M., Berlanga, A.: Optimization of the
  isp parameters of a camera through differential evolution. IEEE Access
  \textbf{8},  143479--143493 (2020)

\bibitem{ignatov2021learned}
Ignatov, A., Chiang, C.M., Kuo, H.K., Sycheva, A., Timofte, R.: Learned
  smartphone isp on mobile npus with deep learning, mobile ai 2021 challenge:
  Report. In: Proceedings of the IEEE/CVF Conference on Computer Vision and
  Pattern Recognition. pp. 2503--2514 (2021)

\bibitem{ignatov2022microisp}
Ignatov, A., Sycheva, A., Timofte, R., Tseng, Y., Xu, Y.S., Yu, P.H., Chiang,
  C.M., Kuo, H.K., Chen, M.H., Cheng, C.M., et~al.: Microisp: processing 32mp
  photos on mobile devices with deep learning. In: European Conference on
  Computer Vision. pp. 729--746. Springer (2022)

\bibitem{ignatov2022learned}
Ignatov, A., Timofte, R., Liu, S., Feng, C., Bai, F., Wang, X., Lei, L., Yi,
  Z., Xiang, Y., Liu, Z., et~al.: Learned smartphone isp on mobile gpus with
  deep learning, mobile ai \& aim 2022 challenge: report. In: European
  Conference on Computer Vision. pp. 44--70. Springer (2022)

\bibitem{ignatov2020replacing}
Ignatov, A., Van~Gool, L., Timofte, R.: Replacing mobile camera isp with a
  single deep learning model. In: Proceedings of the IEEE/CVF Conference on
  Computer Vision and Pattern Recognition Workshops. pp. 536--537 (2020)

\bibitem{johnson2016perceptual}
Johnson, J., Alahi, A., Fei-Fei, L.: Perceptual losses for real-time style
  transfer and super-resolution. In: Computer Vision--ECCV 2016: 14th European
  Conference, Amsterdam, The Netherlands, October 11-14, 2016, Proceedings,
  Part II 14. pp. 694--711. Springer (2016)

\bibitem{li2023spatially}
Li, J., Zhang, Z., Liu, X., Feng, C., Wang, X., Lei, L., Zuo, W.: Spatially
  adaptive self-supervised learning for real-world image denoising. In:
  Proceedings of the IEEE/CVF Conference on Computer Vision and Pattern
  Recognition. pp. 9914--9924 (2023)

\bibitem{liang2021cameranet}
Liang, Z., Cai, J., Cao, Z., Zhang, L.: Cameranet: A two-stage framework for
  effective camera isp learning. IEEE Transactions on Image Processing
  \textbf{30},  2248--2262 (2021)

\bibitem{liu2022deep}
Liu, S., Feng, C., Wang, X., Wang, H., Zhu, R., Li, Y., Lei, L.: Deep-flexisp:
  A three-stage framework for night photography rendering. In: Proceedings of
  the IEEE/CVF Conference on Computer Vision and Pattern Recognition. pp.
  1211--1220 (2022)

\bibitem{liu2023improving}
Liu, W., Li, W., Zhu, J., Cui, M., Xie, X., Zhang, L.: Improving nighttime
  driving-scene segmentation via dual image-adaptive learnable filters. IEEE
  Transactions on Circuits and Systems for Video Technology  (2023)

\bibitem{loshchilov2016sgdr}
Loshchilov, I., Hutter, F.: Sgdr: Stochastic gradient descent with warm
  restarts. arXiv preprint arXiv:1608.03983  (2016)

\bibitem{loshchilov2017decoupled}
Loshchilov, I., Hutter, F.: Decoupled weight decay regularization. arXiv
  preprint arXiv:1711.05101  (2017)

\bibitem{moran2020deeplpf}
Moran, S., Marza, P., McDonagh, S., Parisot, S., Slabaugh, G.: Deeplpf: Deep
  local parametric filters for image enhancement. In: Proceedings of the
  IEEE/CVF conference on computer vision and pattern recognition. pp.
  12826--12835 (2020)

\bibitem{mosleh2020hardware}
Mosleh, A., Sharma, A., Onzon, E., Mannan, F., Robidoux, N., Heide, F.:
  Hardware-in-the-loop end-to-end optimization of camera image processing
  pipelines. In: Proceedings of the IEEE/CVF Conference on Computer Vision and
  Pattern Recognition. pp. 7529--7538 (2020)

\bibitem{onzon2021neural}
Onzon, E., Mannan, F., Heide, F.: Neural auto-exposure for high-dynamic range
  object detection. In: Proceedings of the IEEE/CVF Conference on Computer
  Vision and Pattern Recognition. pp. 7710--7720 (2021)

\bibitem{otsuka2023self}
Otsuka, J., Yoshimura, M., Ohashi, T.: Self-supervised reversed image signal
  processing via reference-guided dynamic parameter selection. arXiv preprint
  arXiv:2303.13916  (2023)

\bibitem{ouyang2023rsfnet}
Ouyang, W., Dong, Y., Kang, X., Ren, P., Xu, X., Xie, X.: Rsfnet: A white-box
  image retouching approach using region-specific color filters. In:
  Proceedings of the IEEE/CVF International Conference on Computer Vision. pp.
  12160--12169 (2023)

\bibitem{paszke2019pytorch}
Paszke, A., Gross, S., Massa, F., Lerer, A., Bradbury, J., Chanan, G., Killeen,
  T., Lin, Z., Gimelshein, N., Antiga, L., et~al.: Pytorch: An imperative
  style, high-performance deep learning library. Advances in neural information
  processing systems  \textbf{32} (2019)

\bibitem{pavithra2021automatic}
Pavithra, G., Radhesh, B.: Automatic image quality tuning framework for
  optimization of isp parameters based on multi-stage optimization approach.
  Electronic Imaging  \textbf{2021}(9),  197--1 (2021)

\bibitem{punnappurath2022day}
Punnappurath, A., Abuolaim, A., Abdelhamed, A., Levinshtein, A., Brown, M.S.:
  Day-to-night image synthesis for training nighttime neural isps. In:
  Proceedings of the IEEE/CVF Conference on Computer Vision and Pattern
  Recognition. pp. 10769--10778 (2022)

\bibitem{qin2022attention}
Qin, H., Han, L., Wang, J., Zhang, C., Li, Y., Li, B., Hu, W.: Attention-aware
  learning for hyperparameter prediction in image processing pipelines. In:
  European Conference on Computer Vision. pp. 271--287. Springer (2022)

\bibitem{qin2023learning}
Qin, H., Han, L., Xiong, W., Wang, J., Ma, W., Li, B., Hu, W.: Learning to
  exploit the sequence-specific prior knowledge for image processing pipelines
  optimization. In: Proceedings of the IEEE/CVF Conference on Computer Vision
  and Pattern Recognition. pp. 22314--22323 (2023)

\bibitem{rombach2022high}
Rombach, R., Blattmann, A., Lorenz, D., Esser, P., Ommer, B.: High-resolution
  image synthesis with latent diffusion models. In: Proceedings of the IEEE/CVF
  conference on computer vision and pattern recognition. pp. 10684--10695
  (2022)

\bibitem{ronneberger2015u}
Ronneberger, O., Fischer, P., Brox, T.: U-net: Convolutional networks for
  biomedical image segmentation. In: Medical Image Computing and
  Computer-Assisted Intervention--MICCAI 2015: 18th International Conference,
  Munich, Germany, October 5-9, 2015, Proceedings, Part III 18. pp. 234--241.
  Springer (2015)

\bibitem{shen2023adaptive}
Shen, H., Zhao, Z.Q., Zhang, W.: Adaptive dynamic filtering network for image
  denoising. In: Proceedings of the AAAI Conference on Artificial Intelligence.
  vol.~37, pp. 2227--2235 (2023)

\bibitem{shi2016real}
Shi, W., Caballero, J., Husz{\'a}r, F., Totz, J., Aitken, A.P., Bishop, R.,
  Rueckert, D., Wang, Z.: Real-time single image and video super-resolution
  using an efficient sub-pixel convolutional neural network. In: Proceedings of
  the IEEE conference on computer vision and pattern recognition. pp.
  1874--1883 (2016)

\bibitem{shibata2016gradient}
Shibata, T., Tanaka, M., Okutomi, M.: Gradient-domain image reconstruction
  framework with intensity-range and base-structure constraints. In:
  Proceedings of the IEEE conference on computer vision and pattern
  recognition. pp. 2745--2753 (2016)

\bibitem{shutova2023ntire}
Shutova, A., Ershov, E., Perevozchikov, G., Ermakov, I., Bani{\'c}, N.,
  Timofte, R., Collins, R., Efimova, M., Terekhin, A., Zini, S., et~al.: Ntire
  2023 challenge on night photography rendering. In: Proceedings of the
  IEEE/CVF Conference on Computer Vision and Pattern Recognition. pp.
  1981--1992 (2023)

\bibitem{simonyan2014very}
Simonyan, K., Zisserman, A.: Very deep convolutional networks for large-scale
  image recognition. arXiv preprint arXiv:1409.1556  (2014)

\bibitem{stevens1957psychophysical}
Stevens, S.S.: On the psychophysical law. Psychological review  \textbf{64}(3),
   153--181 (1957)

\bibitem{tseng2019hyperparameter}
Tseng, E., Yu, F., Yang, Y., Mannan, F., Arnaud, K.S., Nowrouzezahrai, D.,
  Lalonde, J.F., Heide, F.: Hyperparameter optimization in black-box image
  processing using differentiable proxies. ACM Trans. Graph.  \textbf{38}(4),
  27--1 (2019)

\bibitem{van2007edge}
Van De~Weijer, J., Gevers, T., Gijsenij, A.: Edge-based color constancy. IEEE
  Transactions on image processing  \textbf{16}(9),  2207--2214 (2007)

\bibitem{wang2019underexposed}
Wang, R., Zhang, Q., Fu, C.W., Shen, X., Zheng, W.S., Jia, J.: Underexposed
  photo enhancement using deep illumination estimation. In: Proceedings of the
  IEEE/CVF conference on computer vision and pattern recognition. pp.
  6849--6857 (2019)

\bibitem{wu2019visionisp}
Wu, C.T., Isikdogan, L.F., Rao, S., Nayak, B., Gerasimow, T., Sutic, A.,
  Ain-kedem, L., Michael, G.: Visionisp: Repurposing the image signal processor
  for computer vision applications. In: 2019 IEEE International Conference on
  Image Processing (ICIP). pp. 4624--4628. IEEE (2019)

\bibitem{yang2022adaint}
Yang, C., Jin, M., Jia, X., Xu, Y., Chen, Y.: Adaint: Learning adaptive
  intervals for 3d lookup tables on real-time image enhancement. In:
  Proceedings of the IEEE/CVF Conference on Computer Vision and Pattern
  Recognition. pp. 17522--17531 (2022)

\bibitem{yang2022seplut}
Yang, C., Jin, M., Xu, Y., Zhang, R., Chen, Y., Liu, H.: Seplut: Separable
  image-adaptive lookup tables for real-time image enhancement. In: European
  Conference on Computer Vision. pp. 201--217. Springer (2022)

\bibitem{yoshimura2023dynamicisp}
Yoshimura, M., Otsuka, J., Irie, A., Ohashi, T.: Dynamicisp: dynamically
  controlled image signal processor for image recognition. In: Proceedings of
  the IEEE/CVF International Conference on Computer Vision. pp. 12866--12876
  (2023)

\bibitem{yoshimura2023rawgment}
Yoshimura, M., Otsuka, J., Irie, A., Ohashi, T.: Rawgment: noise-accounted raw
  augmentation enables recognition in a wide variety of environments. In:
  Proceedings of the IEEE/CVF Conference on Computer Vision and Pattern
  Recognition. pp. 14007--14017 (2023)

\bibitem{zamir2022restormer}
Zamir, S.W., Arora, A., Khan, S., Hayat, M., Khan, F.S., Yang, M.H.: Restormer:
  Efficient transformer for high-resolution image restoration. In: Proceedings
  of the IEEE/CVF conference on computer vision and pattern recognition. pp.
  5728--5739 (2022)

\bibitem{zeng2020learning}
Zeng, H., Cai, J., Li, L., Cao, Z., Zhang, L.: Learning image-adaptive 3d
  lookup tables for high performance photo enhancement in real-time. IEEE
  Transactions on Pattern Analysis and Machine Intelligence  \textbf{44}(4),
  2058--2073 (2020)

\bibitem{zhang2024lookup}
Zhang, F., Tian, M., Li, Z., Xu, B., Lu, Q., Gao, C., Sang, N.: Lookup table
  meets local laplacian filter: Pyramid reconstruction network for tone
  mapping. Advances in Neural Information Processing Systems  \textbf{36}
  (2024)

\bibitem{zhang2022clut}
Zhang, F., Zeng, H., Zhang, T., Zhang, L.: Clut-net: Learning adaptively
  compressed representations of 3dluts for lightweight image enhancement. In:
  Proceedings of the 30th ACM International Conference on Multimedia. pp.
  6493--6501 (2022)

\bibitem{zhang2021learning}
Zhang, Z., Wang, H., Liu, M., Wang, R., Zhang, J., Zuo, W.: Learning
  raw-to-srgb mappings with inaccurately aligned supervision. In: Proceedings
  of the IEEE/CVF International Conference on Computer Vision. pp. 4348--4358
  (2021)

\bibitem{zhou2021decoupled}
Zhou, J., Jampani, V., Pi, Z., Liu, Q., Yang, M.H.: Decoupled dynamic filter
  networks. In: Proceedings of the IEEE/CVF Conference on Computer Vision and
  Pattern Recognition. pp. 6647--6656 (2021)

\bibitem{zini2023back}
Zini, S., Rota, C., Buzzelli, M., Bianco, S., Schettini, R.: Back to the
  future: a night photography rendering isp without deep learning. In:
  Proceedings of the IEEE/CVF Conference on Computer Vision and Pattern
  Recognition. pp. 1465--1473 (2023)

\bibitem{zou2023iterative}
Zou, Y., Yan, C., Fu, Y.: Iterative denoiser and noise estimator for
  self-supervised image denoising. In: Proceedings of the IEEE/CVF
  International Conference on Computer Vision. pp. 13265--13274 (2023)

\end{thebibliography}
 
\end{document}